\documentclass[12pt]{article}
\usepackage{amssymb,amsmath,amsthm,amsfonts,amscd}
 \usepackage{graphicx}
\newcommand\be{\begin{equation}}
\newcommand\ee{\end{equation}}
\newcommand\bea{\begin{eqnarray}}
\newcommand\eea{\end{eqnarray}}

\newcommand{\fatalpha}{{\bf \alpha \kern -0.44em \alpha}}
\newcommand{\fatsigma}{{\bf \sigma \kern -0.54em \sigma}}
\newcommand{\tpchi}{{\bf D \kern -0.35em D}}
\newcommand{\llambda}{{\bf \lambda \kern -0.45em \lambda}}

\renewcommand{\theequation}{\arabic{equation}}
\renewcommand{\theequation}{\thesection.\arabic{equation}}
\bibliography{plain}
\title{\bf Constructing Entanglement Witness Via Real Skew-Symmetric Operators}\vspace{20mm}
\author{ M. A. Jafarizadeh $^{a,b,c}$
 \thanks{E-mail:jafarizadeh@tabrizu.ac.ir}  ,
 N. Behzadi $^{a,b}$
 \thanks{E-mail:behzadi@tabrizu.ac.ir}
 \\ $^a${\small Department of Theoretical Physics and Astrophysics,
Tabriz University, Tabriz 51664, Iran.} \\ $^b${\small Institute
for Studies in Theoretical Physics and Mathematics, Tehran
19395-1795, Iran.} \\ $^c${\small Research Institute for
Fundamental Sciences, Tabriz 51664, Iran. }} \pagebreak

\vspace{20mm}
\begin{document}
\maketitle \vspace{15mm}
\newpage
\begin{abstract}
In this work, new types of EWs are introduced. They are
constructed by using real skew-symmetric operators defined on a single
party subsystem of a bipartite $d\otimes d$ system and a maximal
entangled state in that system. A canonical form for these
witnesses is proposed which is called canonical EW in corresponding to canonical real skew-symmetric operator. Also for each
possible partition of the canonical real skew-symmetric operator
corresponding EW is obtained. The method used for
$d\otimes d$ case is extended to $d_{1}\otimes d_{2}$ systems.
It is shown that there exist $C^{d_{2}}_{d_{1}}$ distinct
possibilities to construct EWs for a given $d_{1}\otimes
d_{2}$ Hilbert space. The optimality and nd-optimality problem is
studied for each type of EWs. In each step, a large
class of quantum PPT states is introduced. It is shown that among
them there exist entangled PPT states which are
detected by the constructed witnesses. Also the idea of canonical EWs is extended to obtain other EWs with greater PPT entanglement detection power.
\end{abstract}



\vspace{70mm}
\newpage
\section{Introduction}
Entanglement is one of the essential features of quantum physics which has no analogous in classical one.
Entanglement lies in the heart of the quantum information and
quantum computation. It is used as a physical resource which
allows to realize various quantum information and quantum
computation tasks such as quantum cryptography, teleportation,
dense coding, and key distribution
\cite{nielsen1,ekert1,preskill}. The most important problem in
quantum entanglement is determining boundary of the separable
states and entangled ones, which is still not well characterized.
Although the famous Peres-Horodecki criterion based on positive
partial transpose (PPT) explicitly determines this boundary for
low dimensional bipartite systems such as $2\otimes2$ and
$2\otimes3$ \cite{peres,phorodecki}, but
 it has no efficiency for entangled PPT states which appear in the
 higher-dimensional compound quantum systems. On the other hand,
 the entangled PPT
 states belong to the group of entangled states (bound entangled states) which can not be distilled. Therefore, distinguishing these states from
 entangled states that can be distilled (free entangled states) is
 of great importance in quantum communication theory \cite{mprhorodecki1}.
\par The most general approach to solve and characterize the separability problem in quantum theory for any
higher-dimensional physical systems is based upon the notion of
entanglement witness (EW) \cite{mprhorodecki,terhal}. The EWs are
fundamental tool in entanglement theory since it has been shown
that, for any entangled state, there exists at least one EW which
detects its entanglement \cite{mprhorodecki,woronowicz}.
 A Hermitian operator W is said
to be an EW if and only if for all separable states $\rho_{sep}$,
$\mathrm{Tr}(\mathrm{W}\rho_{sep})\geq0$ and at least for one
entangled state $\rho_{ent}$,
$\mathrm{Tr}(\mathrm{W}\rho_{ent})<0$ ( one says that $\rho_{ent}$
is detected by W). Clearly, the construction of EWs is a hard task,
although it is easy to construct an operator W which has negative
expectation value with an entangled state, but it is very
difficult to check that $\mathrm{Tr}(\mathrm{W}\rho)\geq0$ for all
separable states $\rho$ \cite{jafari1,jafari2,jafari3,jafari4}.
There are two types of EWs: decomposable EW (d-EW) that can not
detect any entangled PPT states and non-decomposable one (nd-EW)
which can detect at least one entangled PPT state. It turns out
that the development of appropriate separability criteria reduces
to construction of nd-EWs to detect various entangled PPT states.
There have been several considerable efforts in constructing nd-EWs
(analytically and numerically) for higher-dimensional systems (see
e.g.
\cite{lewenstein1,doherty1,doherty2,sixia,heinz,jafari5,dariusz,bertlman}).
\par In this paper, we describe a method to construct nd-EWs which operate on a bipartite
Hilbert space with arbitrary dimension. Though Breuer attempted in
\cite{heinz} to build such witnesses on the basis of time reversal
transformation but his work was restricted to those physical
systems which had angular momentum symmetry. The approach proposed
in this paper is general and practical for every physical system.
It is based on a skew-symmetric operator which acts on a
d-dimensional single party subsystem and a maximal entangled state
which are called characteristic elements of constructed witness.
In matrix representation, as will be seen, any skew-symmetric
matrix U satisfies the relation
$\langle\alpha^{\ast}|\mathrm{U}|\alpha\rangle=0$ (for every state
vector $|\alpha\rangle$ that lives in the single party subsystem).
It is shown that, by using this property of skew-symmetric
matrices, the related EWs can be constructed analytically for any
$d\otimes d$ systems. From the theory of matrices
\cite{macduffee}, the rank of any skew-symmetric matrix is equal
to an even number $(2n)$. Also every skew-symmetric matrix U can
be written as $\mathrm{U}=\mathrm{Q}\mathrm{J}\mathrm{Q}^{t}$ in
which $\mathrm{J}$ is the canonical form of U (or canonical
skew-symmetric matrix) and Q is an orthogonal matrix
($\mathrm{Q}\mathrm{Q}^{t}=\mathrm{Q}^{t}\mathrm{Q}=\mathrm{I}_{d}$).
A witness which is established by J is called canonical EW. It is
shown that a large number of witnesses are obtained by doing
orthogonal transformations specified by Q, on the characteristic
elements of canonical EW. Thus one is interested to study and
analyze the canonical EWs in details.
\par To develop the idea of the canonical
EW for various ranks that J can achieve, the related canonical EW
is derived. It is proved that for all possible values of
$rank(J)$, ($rank(J)=0,...2n=d$), the corresponding canonical EWs
is optimal. It is also shown that for full-rank J, the related EW is
optimal nd-EW and, when the $rank(J)<4$, it is optimal d-EW. On the
other hand, we introduce a new class of PPT states among which there exist entangled PPT states detected by the above mentioned
witnesses. By the Jamiolkowski isomorphism \cite{jamiol} between
operators and maps, the positive maps corresponding to canonical
EWs are obtained. In the next step, it is assumed that J is
full-rank (that is $d=2n$). In this case, J is partitioned as a
direct sum of block-diagonal matrices, which corresponds to a
partition of n. As illustrated , all of witnesses obtained
according to possible partition of n are optimal EWs.
Also in this step for every partition of n, we construct
entangled PPT state detected by the nd-EW which is given in the
same partition. It is shown that our method can be extended to the
bipartite $d_{1}\otimes d_{2}$ systems. The shape of canonical EWs
are similar to the those ones obtained for $d\otimes d$ systems.
In this case, PPT states are also introduced and by the constructed
canonical EWs, it is shown that some of them are entangled.
\par The paper is organized as follows: In section 2,
we investigate our approach to construct EWs for $d\otimes d$
systems. The canonical EWs along with the PPT states are
introduced. Optimality and nd-optimality for canonical EWs are
discussed and also the positive maps corresponding to canonical
EWs are obtained. In section 3, the canonical EWs corresponding to
possible partition of J and also the related PPT states are
discussed. Section 4 is devoted to describe the canonical EWs and
corresponding PPT states for $d_{1}\otimes d_{2}$ systems. Also in this section, the idea of canonical EW is extended to obtain other EWs. In the end, we summarize our result and present our conclusions.

\section{ Entanglement Witnesses For $d\otimes d$ Systems}
In this section, we are going to describe a method for constructing EWs. Let us consider $\mathcal{H}$ as
a d-dimensional Hilbert space devoted to a single particle
subsystem. U is defined as a real skew-symmetric operator
($\mathrm{U}^{T}=-\mathrm{U}$) which acts on the $\mathcal{H}$. U in
matrix form is
\begin{equation}\label{eq1}
    \mathrm{U}=\sum_{_{i\leq j}}^{d-1}a_{_{ij}}(|i\rangle\langle
    j|-|j\rangle\langle i|)  \qquad\  i,j=0,...,d-1.
    \end{equation}
It is easy to see that for every state $|\alpha\rangle\in \mathcal{H}$ and for every real skew-symmetric matrix U the following relation is satisfied
\begin{equation}\label{eq3}
\langle \alpha^{\ast}|\mathrm{U}|\alpha\rangle=0,
\end{equation} i.e., $|\alpha^{\ast}\rangle$ and $\mathrm{U}|\alpha\rangle$ are orthogonal
to each other. Now let us introduce the following Hermitian operator
\begin{equation}\label{eq4}
    \mathrm{W}=\mathrm{I}_{d}\otimes \mathrm{I}_{d}-d|\psi\rangle\langle\psi|-d(\mathrm{U}^{T}\otimes \mathrm{I}_{d})|\psi\rangle\langle\psi|^{T_{A}}(\mathrm{U}\otimes \mathrm{I}_{d}),
    \end{equation} where  $\mathrm{I}_{d}\otimes \mathrm{I}_{d}$ is the identity operator, T is transposition and $|\psi\rangle$, as defined below, is the $d\otimes d$ maximal entangled
    Bell-state
\begin{equation}\label{eq5}
|\psi\rangle=\frac{1}{\sqrt{d}}\sum_{_{i=0}}^{d-1}|ii\rangle
,\quad\ i=0,...,d-1.
\end{equation}
W has the following expectation value with an arbitrary product state
$|\eta\rangle\otimes|\zeta\rangle$
\begin{equation}\label{eq6}
\langle\eta|\otimes\langle\zeta|\mathrm{W}|\eta\rangle\otimes|\zeta\rangle=1-|\langle\zeta|\eta^{\ast}\rangle|^{2}-|\langle\zeta|\mathrm{U}|\eta\rangle|^{2},
\end{equation}
in which $|\eta^{\ast}\rangle$ and $\mathrm{U}|\eta\rangle$ are orthogonal to each other. It is assumed  $\langle\eta|\mathrm{U}^{T}\mathrm{U}|\eta\rangle\leq1$. If $\mathrm{U}|\eta\rangle$ is normalized then $|\zeta\rangle$, in general, can be written as $|\zeta\rangle=\alpha|\eta^{\ast}\rangle+\beta\mathrm{U}|\eta\rangle$
with $|\alpha|^{2}+|\beta|^{2}=1$ otherwise it can not. Consequently, in both cases, the
expectation values of the W with respect to all separable states
are positive, so the Hermitian operator W can be considered as an
EW. We will show that
this type of EWs have ability to detect entangled PPT states, so
they are nd-EWs. It should be noted that, if we omit the last part
of the EW in (\ref{eq4}), the remainder part is the reduction EW
\cite{jafari1}; in other words,
\begin{equation}\label{eq7}
\mathrm{W}=\mathrm{W}_{red}-d(\mathrm{\mathrm{U}}^{T}\otimes
\mathrm{I}_{d})|\psi\rangle\langle\psi|^{T_{\mathrm{A}}}(\mathrm{\mathrm{U}}\otimes
\mathrm{I}_{d}),
\end{equation}
where
\begin{equation}\label{eq8}
\mathrm{W_{red}}=\mathrm{I}_{d}\otimes
\mathrm{I}_{d}-d|\psi\rangle\langle\psi|.
\end{equation}
\subsection{Canonical Entanglement Witnesses}
In this subsection, we introduce canonical EW and investigate how other EWs are obtained from it.
To illustrate, we know that, from the theory of matrices
\cite{macduffee}, every real skew-symmetric matrix on the d-dimensional
Hilbert space $\mathcal{H}$ can be decomposed as
\begin{equation}\label{eq9}
\mathrm{U}=\mathrm{QJQ}^{T},
\end{equation}
where Q is an orthogonal matrix, i.e. $\mathrm{QQ}^{T}=\mathrm{Q}^{T}\mathrm{Q}=\mathrm{I_{d}}$, and J is a
block-diagonal one which is called the canonical form of U
\begin{equation}\label{eq10}
\mathrm{J}=\mathrm{j}_{0}\oplus \mathrm{j}_{1}\oplus \mathrm{j}_{2}\oplus...\oplus \mathrm{j}_{n-1},
\end{equation}
where
\begin{equation}\label{eq11}
\mathrm{j}_{i}=\left(%
\begin{array}{cc}
0 & \lambda_{_{i}} \\
-\lambda_{_{i}}&0 \\
\end{array}%
\right).
\end{equation}
The scalars $\lambda_{0}$, $\lambda_{1}$,$\cdots$,$\lambda_{n-1}$,
which appear in the $\mathrm{J}$, are invariant factors of
$\mathrm{U}$ (they are invariant under orthogonal
transformations). It is clear that the rank of every
real skew-symmetric matrix $\mathrm{U}$ is always an even number
($2n$); therefore, if the matrix U is full-rank, then $2n=d$ and if
d is an odd number, U can not be full-rank anywise. Also it is clear that the
eigenvalues of U are complete imaginary or zero, but, if U is
full-rank, then all of its eigenvalues will be imaginary. In
addition, the condition
$\langle\eta|\mathrm{U}^{T}\mathrm{U}|\eta\rangle\leq1$ yields
$\mathrm{J}\mathrm{J}^{T}=\mathrm{J}^{T}\mathrm{J}\leq
\mathrm{I_{d}}$ which leads to a condition on $\lambda_{i}$s, i.e. $\lambda_{i}\leq1$ for $i=0,\cdots, n-1$. Now we are ready to
introduce canonical EW. Consider the following Hermitian operator
\begin{equation}\label{eq12}
\mathrm{W}_{C}=\mathrm{I}_{d}\otimes
\mathrm{I}_{d}-d|\psi\rangle\langle\psi|-d(\mathrm{J}^{T}\otimes
\mathrm{I}_{d})|\psi\rangle\langle\psi|^{T_{A}}(\mathrm{J}\otimes
\mathrm{I}_{d})
\end{equation}
By the same prescription proposed for W in equation (\ref{eq4}) to be
as an EW, $\mathrm{W}_{C}$ can also be an EW. It is based on two
characteristic elements: J and $|\psi\rangle\langle\psi|$. If we
do some transformations on these operators, we obtain other EWs.
To further illustrate, let $|\psi\rangle\langle\psi|$ be
transformed to $(\mathrm{Q}^{T}\otimes
\mathrm{I}_{d})|\psi\rangle\langle\psi|(\mathrm{Q}\otimes
\mathrm{I}_{d})$ then $\mathrm{W}_{C}$ becomes as
\begin{equation}\label{eq13}
\mathrm{W}_{\psi}=\mathrm{I}_{d}\otimes
\mathrm{I}_{d}-d(\mathrm{Q}^{T}\otimes
\mathrm{I}_{d})|\psi\rangle\langle\psi|(\mathrm{Q}\otimes
\mathrm{I}_{d})-d(\mathrm{J}^{T}\mathrm{\mathrm{Q}}^{T}\otimes
\mathrm{I}_{d})|\psi\rangle\langle\psi|^{T_{A}}(\mathrm{QJ}\otimes
\mathrm{I}_{d})
\end{equation}
where Q can be any orthogonal matrix corresponding to any orthogonal
transformation in the d-dimensional Hilbert space. The subscript
$|\psi\rangle$ denotes the class of operators obtained by any local
orthogonal transformation $\mathrm{Q}\otimes \mathrm{I}_{d}$ which
is preformed on $|\psi\rangle\langle\psi|$. Now the calculation of the
expectation values of $\mathrm{W}_{\psi}$ over all separable
states yields the following equation
\begin{equation}\label{eq14}
\langle\eta|\otimes\langle\zeta|\mathrm{W}_{\psi}|\eta\rangle\otimes|\zeta\rangle=1-|\langle\zeta|\mathrm{Q}|\eta^{\ast}\rangle|^{2}-|\langle\zeta|\mathrm{QJ}|\eta\rangle|^{2}.
\end{equation}
By the same argument sketched in equation (\ref{eq6}), the above mentioned expectation values are always positive.
Therefore, the class of operators, $\mathrm{W}_{\psi}$, can be
considered as EWs.
The other class is obtained by transforming J in to $\mathrm{Q}\mathrm{J}{\mathrm{Q}}^{T}$ as
\begin{equation}\label{eq15}
\mathrm{W}_{J}=\mathrm{I}_{d}\otimes
\mathrm{I}_{d}-d|\psi\rangle\langle\psi|-d(\mathrm{Q}\mathrm{J}^{T}\mathrm{\mathrm{Q}}^{T}\otimes
\mathrm{I}_{d})|\psi\rangle\langle\psi|^{T_{A}}(\mathrm{Q}\mathrm{J}\mathrm{Q}^{T}\otimes
\mathrm{I}_{d}),
\end{equation}where the subscript J denotes the class of all operators obtained
by any orthogonal transformation Q which is done on J. By the
equation (\ref{eq9}), the operator in (\ref{eq15}) is the EW denoted in (\ref{eq4}).
On the other hand, we can see that the class of EWs $\mathrm{W}_{\psi}$ and $\mathrm{W}_{J}$ are related to
each other by local orthogonal transformation $(\mathrm{Q}\otimes
\mathrm{I}_{d})$, i.e.
\begin{equation}\label{eq16}
\mathrm{W}_{\psi}=(\mathrm{Q}^{T}\otimes
\mathrm{I}_{d})\mathrm{W}_{J}(\mathrm{Q}\otimes \mathrm{I}_{d}),
\end{equation}
therefore they are locally equivalent. These results strongly
motivate us to go to study canonical EW, $\mathrm{W}_{C}$, in
detail. Hence the rest of the paper will be devoted to describe
the properties of the canonical EW. To this end, we see that the action of $\mathrm{J}$ on
the basis states of the d-dimensional single party Hilbert space $\mathcal{H}$ is as
\begin{equation}\label{eq17}
\mathrm{J}|2i\rangle=-|2i+1\rangle ,\qquad\
\mathrm{J}|2i+1\rangle=|2i\rangle, \qquad\
i=0,...,n-1,
\end{equation}
and
\begin{equation}\label{eq18}
\mathrm{J}|i\rangle=0 ,\qquad\ i=2n,...,d-1.
\end{equation}
By referring to equation (\ref{eq12}) and using the equations (\ref{eq17}) and (\ref{eq18}), $\mathrm{W}_{C}$ is obtained in expanded form as$$\mathrm{W}_{C}=\sum_{_{i=0}}^{n-1}(1-\lambda_{i}^{2})(|2i,2i+1\rangle\langle2i,2i+1|+|2i+1,2i\rangle\langle2i+1,2i|$$
$$-|2i,2i\rangle\langle2i+1,2i+1|-|2i+1,2i+1\rangle\langle2i,2i|)$$
$$+\sum_{_{i\neq j=0}}^{n-1}(|2i,2j\rangle\langle2i,2j|
+|2i,2j+1\rangle\langle2i,2j+1|
+|2i+1,2j\rangle\langle2i+1,2j|$$
$$+|2i+1,2j+1\rangle\langle2i+1,2j+1|-|2i,2i\rangle\langle2j,2j|-|2i,2i\rangle\langle2j+1,2j+1|$$
$$-|2i+1,2i+1\rangle\langle2j,2j|-|2i+1,2i+1\rangle\langle2j+1,2j+1|$$
$$-\lambda_{_{i}}\lambda_{_{j}}(|2i+1,2j\rangle\langle2j+1,2i|-|2i+1,2j+1\rangle\langle2j,2i|$$
$$-|2i,2j\rangle\langle2j+1,2i+1|+|2i,2j+1\rangle\langle2j,2i+1|))$$
$$+\sum_{_{i=0}}^{n-1}\sum_{_{j=2n}}^{d-1}(|2i,j\rangle\langle2i,j|+|2i+1,j\rangle\langle2i+1,j|+|j,2i\rangle\langle
j,2i|+|j,2i+1\rangle\langle j,2i+1|$$
$$-|2i,2i\rangle\langle j,j|-|2i+1,2i+1\rangle\langle
j,j|-|j,j\rangle\langle 2i,2i|-|j,j\rangle\langle 2i+1,2i+1|)$$
\be\label{eq19}+\sum_{_{i,j=2n}}^{d-1}(|i,j\rangle\langle
i,j|-|i,i\rangle\langle j,j|).\ee Consider the following operator
$$\mathrm{O}^{T_{A}}=\sum_{_{i=0}}^{n-1}(1-\lambda_{_{i}}^{2})(|2i,2i+1\rangle\langle2i,2i+1|+|2i+1,2i\rangle\langle2i+1,2i|$$
\be\label{eq20}-|2i,2i\rangle\langle2i+1,2i+1|-|2i+1,2i+1\rangle\langle2i,2i|).\ee
The $\mathrm{W}_{C}$ is composed of the $\mathrm{O}^{t_{A}}$ and
the remainder one which is called
$\mathrm{W}_{C}(\lambda_{0},\lambda_{1},...,\lambda_{n-1})$, i.e.
\begin{eqnarray}\label{eq21}
\mathrm{W}_{C}=\mathrm{O}^{T_{A}}+\mathrm{W}_{C}(\lambda_{0},\lambda_{1},...,\lambda_{n-1}).
\end{eqnarray}
We see that the operator $\mathrm{O}$ is positive; therefore, from
\cite{lewenstein2} for every PPT state $\rho$,
$\mathrm{Tr}(O^{t_{A}}\rho)\geq0$. It becomes clear that the
following inequality is satisfied for every PPT state
\begin{eqnarray}\label{eq22}
\mathrm{Tr}(\mathrm{W}_{C}\rho)\geq
\mathrm{Tr}(\mathrm{W}_{C}(\lambda_{0},\lambda_{1},...,\lambda_{n-1})\rho).
\end{eqnarray}
This inequality enables us to take $\lambda_{i}=1$ for
$i=0,...,n-1$ so $\mathrm{O}^{T_{A}}$ becomes zero and
$\mathrm{W}_{C}=\mathrm{W}_{C}(1,1,...,1)$. From now on, we keep
discussing on $\mathrm{W}_{C}$ for $\lambda_{i}=1$ ($i=0,...,n-1$)
and rewriting $\mathrm{W}_{C}$ as
$$\mathrm{W}_{C}=\sum_{_{i\neq j=0}}^{n-1}(|2i,2j\rangle\langle2i,2j|
+|2i,2j+1\rangle\langle2i,2j+1|
+|2i+1,2j\rangle\langle2i+1,2j|$$
$$+|2i+1,2j+1\rangle\langle2i+1,2j+1|-|2i,2i\rangle\langle2j,2j|-|2i,2i\rangle\langle2j+1,2j+1|$$
$$-|2i+1,2i+1\rangle\langle2j,2j|-|2i+1,2i+1\rangle\langle2j+1,2j+1|$$
$$-|2i+1,2j\rangle\langle2j+1,2i|+|2i+1,2j+1\rangle\langle2j,2i|$$
$$+|2i,2j\rangle\langle2j+1,2i+1|-|2i,2j+1\rangle\langle2j,2i+1|)$$
$$+\sum_{_{i=0}}^{n-1}\sum_{_{j=2n}}^{d-1}(|2i,j\rangle\langle2i,j|+|2i+1,j\rangle\langle2i+1,j|+|j,2i\rangle\langle
j,2i|+|j,2i+1\rangle\langle j,2i+1|$$
$$-|2i,2i\rangle\langle j,j|-|2i+1,2i+1\rangle\langle
j,j|-|j,j\rangle\langle 2i,2i|-|j,j\rangle\langle 2i+1,2i+1|)$$
\be\label{eq23}+\sum_{_{i,j=2n}}^{d-1}(|i,j\rangle\langle
i,j|-|i,i\rangle\langle j,j|).\ee
To disambiguate, $\mathrm{W}_{C}$ can be briefly written as
\begin{eqnarray}\label{eq24}
\mathrm{W}_{C}=\mathrm{O}_{_{1}}^{T_{A}}+\mathrm{O}_{_{2}}^{T_{A}}+\mathrm{W}_{OPC},
\end{eqnarray}
where
$$\mathrm{O}_{_{1}}^{T_{A}}=\sum_{_{i=0}}^{n-1}\sum_{_{j=2n}}^{d-1}(|2i,j\rangle\langle2i,j|+|2i+1,j\rangle\langle2i+1,j|+|j,2i\rangle\langle
j,2i|+|j,2i+1\rangle\langle j,2i+1|$$
\be\label{eq25}-|2i,2i\rangle\langle j,j|-|2i+1,2i+1\rangle\langle
j,j|-|j,j\rangle\langle 2i,2i|-|j,j\rangle\langle 2i+1,2i+1|),\ee
\begin{eqnarray}\label{eq26}
\mathrm{O}_{_{2}}^{T_{A}}=\sum_{_{i,j=2n}}^{d-1}(|i,j\rangle\langle
i,j|-|i,i\rangle\langle j,j|)
\end{eqnarray}
and
$$\mathrm{W}_{OPC}=\sum_{_{i\neq j=0}}^{n-1}(|2i,2j\rangle\langle2i,2j|
+|2i,2j+1\rangle\langle2i,2j+1|
+|2i+1,2j\rangle\langle2i+1,2j|$$
$$+|2i+1,2j+1\rangle\langle2i+1,2j+1|-|2i,2i\rangle\langle2j,2j|-|2i,2i\rangle\langle2j+1,2j+1|$$
$$-|2i+1,2i+1\rangle\langle2j,2j|-|2i+1,2i+1\rangle\langle2j+1,2j+1|$$
$$-|2i+1,2j\rangle\langle2j+1,2i|+|2i+1,2j+1\rangle\langle2j,2i|$$
\be\label{eq27}+|2i,2j\rangle\langle2j+1,2i+1|-|2i,2j+1\rangle\langle2j,2i+1|).\ee
Note that if the rank of
J ($2n$) is two then $\mathrm{W}_{OPC}$ will be zero and therefore by considering the equation (\ref{eq24}), it is concluded that $\mathrm{W}_{C}$ is a d-EW. Since we are interested in dealing with
nd-EWs then those $d\otimes d$ quantum systems for which $rank(J)\geq4$ are discussed. Hence we conclude that, by this approach, nd-EWs can be constructed only for those systems with $d\geq4$. Clearly if J is full-rank ($2n=d$) then in equation (\ref{eq24}), $\mathrm{O}^{T_{A}}_{1}$ and $\mathrm{O}^{T_{A}}_{2}$ will be zero.
Now we claim that the $\mathrm{W}_{C}$ type witnesses are able to
detect entangled PPT states so they are nd-EWs.
\subsection{PPT states}
This subsection is devoted to construct PPT states and determine a
subset of them as a set of entangled PPT states whose
entanglement are detected by the EWs introduced in the previous
subsection. Let us write the following operator
$$\rho=\frac{1}{\mathcal{N}}(a_{0}|\psi\rangle\langle\psi|+a_{0}\sum_{i=0}^{n-1}(|2i,2i\rangle\langle2i,2i|+|2i+1,2i+1\rangle\langle2i+1,2i+1|$$
$$-|2i,2i\rangle\langle2i+1,2i+1|-|2i+1,2i+1\rangle\langle2i,2i|)$$
$$+\sum_{i=0}^{n-1}(a_{2i+2,2i}|2i+2,2i\rangle\langle2i+2,2i|+a_{2i+1,2i+3}|2i+1,2i+3\rangle\langle2i+1,2i+3|$$
$$-C_{i}(|2i+2,2i\rangle\langle2i+1,2i+3|+|2i+1,2i+3\rangle\langle2i+2,2i|))$$
$$+\sum_{i\neq j=0,i-j\neq 1,j-i\neq
n-1}^{n-1}a_{2i,2j}|2i,2j\rangle\langle2i,2j|+\sum_{i\neq
j=0,j-i\neq 1,i-j\neq
n-1}^{n-1}a_{2i+1,2j+1}|2i+1,2j+1\rangle\langle2i+1,2j+1|$$
$$+\sum_{i,j=0}^{n-1}(a_{2i,2j+1}|2i,2j+1\rangle\langle2i,2j+1|+a_{2i+1,2j}|2i+1,2j\rangle\langle2i+1,2j|)$$
$$+\sum_{i=2n}^{d-1}\sum_{j=0}^{n-1}(a_{2j,i}|2j,i\rangle\langle2j,i|+a_{2j+1,i}|2j+1,i\rangle\langle2j+1,i|
+a_{i,2j}|i,2j\rangle\langle i,2j|+a_{i,2j+1}|i,2j+1\rangle\langle
i,2j+1|)$$ \be\label{eq28}+ \sum_{i\neq
j=2n}^{d-1}a_{ij}|i,j\rangle\langle i,j|)
 ,\ee
where $\mathcal{N}$ is the normalization factor and equals to
$$\mathcal{N}=(d+2n)a_{0}+\sum_{i\neq j=0}^{n-1}(a_{_{2i,2j}}+a_{_{2i+1,2j+1}})+\sum_{i,j=0}^{n-1}(a_{_{2i,2j+1}}+a_{_{2i+1,2j}})$$
\be\label{eq29}+\sum_{i=2n}^{d-1}\sum_{j=0}^{n-1}(a_{_{2j,i}}+a_{_{2j+1,i}}+a_{_{i,2j}}+a_{_{i,2j+1}})+\sum_{i\neq
j=2n}^{d-1}a_{_{i,j}}. \ee The positivity conditions impose that
all of the multipliers which appear in the $\rho$ are positive semi
definite and in addition the following inequality must be
satisfied
\begin{eqnarray}\label{eq30}
a_{_{2i+1,2i+3}}a_{_{2i+2,2i}}\geq C_{i}^{2},\quad \quad
i=0,...,n-1,
\end{eqnarray}
where addition in the subscripts is done by module (2n). Also the
$\mathrm{PPT}$ conditions are as follows:
$$a_{2i,2j}\\\\\\\ a_{2j,2i}\geq a_{0}^{2},\quad\quad i,j=0,...,n-1,\quad\quad i\neq
j,$$
$$a_{2i+1,2j+1}\\\\\\\ a_{2j+1,2i+1}\geq a_{0}^{2},\quad \quad i,j=0,...,n-1,\quad\quad i\neq
j,$$
$$a_{2i,2j+1}\\\\\\\ a_{2j+1,2i}\geq a_{0}^{2},\quad \quad i,j=0,...,n-1,\quad\quad i\neq
j,$$
$$a_{2i+1,2i}\\\\\\\ a_{2i+2,2i+3}\geq C_{i}^{2},\quad\quad i=0,...,n-1,$$
$$a_{2j,i}\\\\\\\ a_{i,2j}\geq a_{0}^{2},\quad\quad j=0,...,n-1,\quad\quad i=2n,...,d-1,$$
$$a_{2j+1,i}\\\\\\\ a_{i,2j+1}\geq a_{0}^{2},\quad \quad j=0,...,n-1,\quad\quad i=2n,...,d-1,$$
\be\label{eq31}a_{i,j}\\\\\\\ a_{j,i}\geq a_{0}^{2},\quad \quad
i,j=2n,...,d-1,\quad\quad i\neq j.\ee Therefore, by these two
groups of inequalities, the operator $\rho$ becomes as a density
operator with positive partial transpose. In the next step, it is
shown that the expectation value of the witness $\mathrm{W}_{C}$
with respect to the $\rho$, under positivity and PPT conditions,
really fulfills the following inequality (the proving of the
following Lower bound is given in the appendix A.)
\begin{eqnarray}\label{eq32}
\mathrm{Tr}(\mathrm{W}_{C}\rho)\geq\frac{-2n}{d+4n^{2}+(d-2n)(d+2n-1)},\qquad
 \qquad n=2,3,4,\cdots.
\end{eqnarray}
Hence if the $\mathrm{PPT}$ state $\rho$ satisfies the following
inequality
\begin{eqnarray}\label{eq33}
\frac{-2n}{d+4n^{2}+(d-2n)(d+2n-1)}\leq\mathrm{Tr}(\mathrm{W}_{C}\rho)<0,\qquad
 \qquad n=2,3,4,\cdots,
\end{eqnarray}
then it will be an entangled PPT state whose entanglement is
detected by $\mathrm{W}_{C}$. On the other hand, the witness
$\mathrm{W}_{C}$ which can detect the entangled PPT state $\rho$,
becomes as a nd-EW which proves our claim. It is also seen that
the lower bound of the inequality in (\ref{eq28}) depends on the
rank of the matrix J and the dimension of the Hilbert space of the
single party subsystem. It is also obvious that if J is full-rank,
then the lower bound becomes smaller. Finally; when
the state $\rho$ violate the PPT conditions, the expectation value
of the entanglement witness $\mathrm{W}_{C}$ with respect to the
density operator $\rho$ satisfies the following inequality
$$-\frac{4n(n-1)+(d-2n)(d+2n-1)}{d+2n}\leq\mathrm{Tr}(\mathrm{W}_{C}\rho)<\frac{-2n}{d+4n^{2}+(d-2n)(d+2n-1)},$$
\be\label{34}n=2,3,4,\cdots.\ee.Clearly the lower bound becomes
greater when the matrix J is full-rank.

\subsection{Optimal Canonical Entanglement Witnesses}

Now we discuss at first the optimality of canonical EWs by investigating optimality of canonical EW
when J is full-rank (2n=d). The proving of optimality for the
others ($2n<d$) is similar to this one. Secondly, we describe nd-optimality of canonical EWs. There exist different
definitions of optimal entanglement witness. Our description is
based on the definition introduced by Lewenstein et.al., \cite{lewenstein2}. One has two EWs $\mathrm{W}_{1}$ and
$\mathrm{W}_{2}$, $\mathrm{W}_{2}$ is finer than $\mathrm{W}_{1}$
if they differ by a positive operator P. We say that W is optimal
iff for all P and $\epsilon>0$,
$\mathrm{W'}=(1+\epsilon)\mathrm{W}-\epsilon \mathrm{P}$ is not an
EW. P is positive operator and
$\mathrm{P}\mathrm{P}_{\mathrm{W}}=0$ where
$\mathrm{P}_{\mathrm{W}}=\{|\gamma\rangle,
\mathrm{Tr}(\mathrm{W}|\gamma\rangle\langle\gamma|=0)\}$ in which
$|\gamma\rangle$ is separable state. Since any positive operator can be written as a convex combination of pure product states so let us assume that
$\mathrm{P}=|\psi\rangle\langle\psi|$ in which
\begin{eqnarray}\label{eq35}
|\psi\rangle=\sum_{i,j=0}^{d-1}a_{i,j}|i,j\rangle.
\end{eqnarray}
By referring to the equation (\ref{eq6}) and using J instead of U
, a typical separable state $|\gamma\rangle\in
\mathrm{P}_{\mathrm{W}_{C}}$ has the following form
\begin{eqnarray}\label{eq36}
|\gamma\rangle=|\eta\rangle\otimes(\alpha|\eta^{\ast}\rangle+\beta
J|\eta\rangle),
\end{eqnarray}
where $|\alpha|^{2}+|\beta|^{2}=1$. We define
$|\eta\rangle=\sum_{i=0}^{n-1}(\eta_{_{2i}}|2i\rangle+\eta_{_{2i+1}}|2i+1\rangle)$
such that
$\sum_{i=0}^{n-1}(|\eta_{_{2i}}|^{2}+|\eta_{_{2i+1}}|^{2})=1$. For
simplicity we choose $\alpha=1$ and $\beta=0$
($|\gamma\rangle=|\eta\rangle\otimes|\eta^{\ast}\rangle$).
Therefore, the separable state $|\gamma\rangle$ in expanded form
is written as
$$|\gamma\rangle=\sum_{i,j=0}^{n-1}\eta_{_{2i}}\eta^{\ast}_{_{2j}}|2i,2j\rangle+\eta_{_{2i}}\eta^{\ast}_{_{2j+1}}|2i,2j+1\rangle$$
\be\label{37}+\eta_{_{2i+1}}\eta^{\ast}_{_{2j}}|2i+1,2j\rangle+\eta_{_{2i+1}}\eta^{\ast}_{_{2j+1}}|2i+1,2j+1\rangle.\ee
It is proven P=0 as:\\
\textbf{1}. \qquad $\eta_{_{2i}}=\delta_{ik}$ \quad\quad   which
gives the following separable state
\begin{eqnarray}\label{eq38}
|\gamma\rangle=|2k,2k\rangle \longrightarrow
\langle\gamma|\psi\rangle=a_{2k,2k}=0.
\end{eqnarray}.
\textbf{2}. \qquad $\eta_{_{2i+1}}=\delta_{ik}$ \quad\quad
\begin{eqnarray}\label{eq39}
|\gamma\rangle=|2k+1,2k+1\rangle \longrightarrow
\langle\gamma|\psi\rangle=a_{2k+1,2k+1}=0.
\end{eqnarray}
\textbf{3}. \qquad $\eta_{_{2k}}=\alpha',\quad\quad
\eta_{_{2k+1}}=\beta'$, \quad\quad$|\alpha'|^{2}+|\beta'|^{2}=1$,
$$|\gamma\rangle=|\alpha'|^{2}|2k,2k\rangle+\alpha'\beta'^{\ast}|2k,2k+1\rangle+\beta'\alpha'^{\ast}|2k+1,2k\rangle+|\beta'|^{2}|2k+1,2k+1\rangle,$$,
$$\langle\gamma|\psi\rangle=\beta'\alpha'^{\ast}a_{2k+1,2k}+\alpha'\beta'^{\ast}a_{2k,2k+1}=0$$
\be\label{40} \longrightarrow a_{2k,2k+1}=0,\quad\quad
a_{2k+1,2k}=0.\ee \textbf{4}. \qquad $\eta_{_{2k}}=\alpha',\quad\quad
\eta_{_{2l}}=\beta'$ ,\quad\quad
$k<l$,\quad\quad$|\alpha'|^{2}+|\beta'|^{2}=1,$
$$|\gamma\rangle=|\alpha'|^{2}|2k,2k\rangle+\alpha'\beta'^{\ast}|2k,2l\rangle+\beta'\alpha'^{\ast}|2l,2k\rangle+|\beta'|^{2}|2l,2l\rangle,$$
$$\langle\gamma|\psi\rangle=\alpha'\beta'^{\ast}a_{2k,2l}+\beta'\alpha^{\ast}a_{2l,2k}=0$$
\be\label{41} \longrightarrow a_{2k,2l}=0,\quad\quad
a_{2l,2k}=0.\ee \textbf{5}. \qquad $\eta_{_{2k+1}}=\alpha',\quad\quad
\eta_{_{2l+1}}=\beta'$, \quad\quad
$k<l$\quad,\quad$|\alpha'|^{2}+|\beta'|^{2}=1,$
$$|\gamma\rangle=|\alpha'|^{2}|2k+1,2k+1\rangle+\alpha'\beta'^{\ast}|2k+1,2l+1\rangle+\beta'\alpha'^{\ast}|2l+1,2k+1\rangle+|\beta'|^{2}|2l+1,2l+1\rangle,$$
$$\langle\gamma|\psi\rangle=\alpha'\beta'^{\ast}a_{2k+1,2l+1}+\beta'\alpha'^{\ast}a_{2l+1,2k+1}=0$$
\be\label{42} \longrightarrow a_{2k+1,2l+1}=0,\quad\quad
a_{2l+1,2k+1}=0.\ee \textbf{6}. \qquad $\eta_{_{2k}}=\alpha',\quad\quad
\eta_{_{2l+1}}=\beta'$ ,\quad\quad
$k<l$,\quad\quad$|\alpha'|^{2}+|\beta'|^{2}=1$
$$|\gamma\rangle=|\alpha'|^{2}|2k,2k\rangle+\alpha'\beta'^{\ast}|2k,2l+1\rangle+\beta'\alpha'^{\ast}|2l+1,2k\rangle+|\beta'|^{2}|2l+1,2l+1\rangle,$$
$$\langle\gamma|\psi\rangle=\alpha'\beta'^{\ast}a_{2k,2l+1}+\beta'\alpha'^{\ast}a_{2l+1,2k}=0$$
\be\label{43} \longrightarrow a_{2k,2l+1}=0,\quad\quad
a_{2l+1,2k}=0.\ee \textbf{7}. \qquad $\eta_{_{2k+1}}=\alpha',\quad\quad
\eta_{_{2l}}=\beta'$ ,\quad\quad
$k<l$,\quad\quad$|\alpha'|^{2}+|\beta'|^{2}=1,$
$$|\gamma\rangle=|\beta'|^{2}|2l,2l\rangle+\beta'\alpha'^{\ast}|2l,2k+1\rangle+\alpha'\beta'^{\ast}|2k+1,2l\rangle+|\alpha'|^{2}|2k+1,2k+1\rangle,$$
$$\langle\gamma|\psi\rangle=\alpha'\beta'^{\ast}a_{2k+1,2l}+\beta'\alpha'^{\ast}a_{2l,2k+1}=0$$
\be\label{44} \longrightarrow a_{2l,2k+1}=0,\quad\quad
a_{2k+1,2l}=0.\ee
These equations explicitly show that P=0 and therefore the
canonical EW is optimal. The proving of optimality for
$\mathrm{W}_{C}$, when J is not full-rank (and specially is zero),
is similar to the previous one except by noting that the typical
separable state $|\gamma\rangle\in \mathrm{P}_{C}$ has the form
$|\gamma\rangle=|\eta\rangle\otimes|\eta^{\ast}\rangle$ where
$|\eta\rangle=\sum_{i=0}^{n-1}(\eta_{_{2i}}|2i\rangle+\eta_{_{2i+1}}|2i+1\rangle)+\sum_{i=2n}^{d-1}\eta_{i}|i\rangle$.
Therefore, it is concluded that the canonical EW (\ref{eq12}) is
optimal for all ranks of J ($rank(\mathrm{J})=0,...,2n=d$), in
other words, its optimality is independent from the rank of J.
On the other hand, to discuss nd-optimality, we define
$d_{\mathrm{W}_{C}}=\{\rho\geq0| \rho^{T_{\mathrm{A}}}\geq0,\quad \mathrm{Tr}(\mathrm{W}_{c}\rho)<0\}$, i.e., the set of
entangled PPT states detected by $\mathrm{W}_{C}$. Given two
nd-EWs $\mathrm{W}_{C1}$ and $\mathrm{W}_{C2}$, we say that
$\mathrm{W}_{C2}$ is nd-finer than $\mathrm{W}_{C1}$, if
$d_{\mathrm{W}_{C1}}\subseteq d_{\mathrm{W}_{C2}}$, i.e., if
all of the entangled PPT states detected by $\mathrm{W}_{C1}$ are
also detected by $\mathrm{W}_{C2}$. We say that $\mathrm{W}_{C}$
is optimal nd-EW, if there exists no other nd-EW which is nd-finer
than it. So by keeping this in mind, we determine optimal nd-EW
among canonical EWs. By referring to equation (\ref{eq24}), we see that the operators $\mathrm{O}_{_{1}}$ and $\mathrm{O}_{_{2}}$
are positive, so $\mathrm{O}^{T_{A}}_{_{1}}$ and $\mathrm{O}^{T_{A}}_{_{2}}$ are optimal d-EWs \cite{lewenstein2} and
$\mathrm{W}_{OPC}$ is a canonical EW on the $2n\otimes2n$ Hilbert
space which is a subspace of $d\otimes d$ one (the proving that $\mathrm{W}_{OPC}$ is an EW, is given in appendix B). We know that
for any PPT state the following inequality is satisfied
\begin{eqnarray}\label{eq45}
\mathrm{Tr}(\mathrm{W}_{C}\rho)\geq
\mathrm{Tr}(\mathrm{W}_{OPC}\rho).
\end{eqnarray}
From the Lewenstein definition of an optimal nd-EW, it is clear
that each entangled PPT state detected by $\mathrm{W}_{C}$, is
also detected by $\mathrm{W}_{OPC}$. It is proven that
$\mathrm{W}_{OPC}$ is an optimal nd-EW. To this aim we say that a
nd-EW, W, is optimal nd-EW iff for all decomposable operator D
($\mathrm{D}=\mathrm{P}+\mathrm{Q}^{t_{\mathrm{A}}} $ where P and
Q are positive operators) and $\epsilon>0$,
$\mathrm{W}'=(1+\epsilon)\mathrm{W}-\epsilon \mathrm{D}$ is not an
EW. In the proving of optimality for canonical EWs, It was shown
that $\mathrm{P}=0$. To illustrate that $\mathrm{W}_{OPC}$ is an
optimal nd-EW, it must be shown that for $\mathrm{Q}^{T_{A}}\mathrm{P}_{\mathrm{W}_{OPC}}=0$ then $\mathrm{Q}^{T_{A}}=0$. Let
us assume that $\mathrm{Q}=|\varphi\rangle\langle\varphi|$ in
which $|\varphi\rangle=\sum_{i,j=0}^{d-1}a_{i,j}|i,j\rangle$. As
previously, since J is full-rank on $2n$-dimensional subspace of the d-dimensional one party Hilbert space $\mathcal{H}$, a
typical separable state $|\gamma\rangle\in \mathrm{P}_{W_{OPC}}$ which lies in the $2n\otimes2n$ subspace (see appendix B),
is written as
$|\vartheta\rangle=|\eta\rangle\otimes(\alpha|\eta^{\ast}\rangle+\beta
\mathrm{J}|\eta\rangle)$. On the other hand, since
$\mathrm{Tr}(\mathrm{Q}^{t_{A}}|\alpha\rangle\langle\alpha|\otimes|\beta\rangle\langle\beta|)=\mathrm{Tr}(\mathrm{Q}|\alpha^{\ast}\rangle\langle\alpha^{\ast}|\otimes|\beta\rangle\langle\beta|)$
then we calculate the expectation values of Q with the
products
$|\vartheta\rangle=|\eta^{\ast}\rangle\otimes(\alpha|\eta^{\ast}\rangle+\beta
\mathrm{J}|\eta\rangle)$. $|\vartheta\rangle$ in expanded form is
$$|\vartheta\rangle=\sum_{i,j=0}^{n-1}\eta^{\ast}_{_{2i}}(\alpha
\eta^{\ast}_{_{2j}}+\beta\eta_{_{2j+1}})|2i,2j\rangle+\eta^{\ast}_{_{2i}}(\alpha
\eta^{\ast}_{_{2j+1}}-\beta\eta_{_{2j}})|2i,2j+1\rangle$$
\be\label{46}+\eta^{\ast}_{_{2i+1}}(\alpha
\eta^{\ast}_{_{2j}}+\beta\eta_{_{2j+1}})|2i+1,2j\rangle+\eta^{\ast}_{_{2i+1}}(\alpha
\eta^{\ast}_{_{2j+1}}-\beta\eta_{_{2j}})|2i+1,2j+1\rangle.\ee
One can show $\mathrm{Q}=0$ on the subspace $2n\otimes2n$ as below\\
\textbf{1}. \qquad $\eta_{_{2i}}=\delta_{ik}$,  $\alpha=1 ,
\beta=0$\quad\quad which gives the following separable state
\begin{eqnarray}\label{47}
|\vartheta\rangle=|2k,2k\rangle \longrightarrow
\langle\vartheta|\varphi\rangle=a_{2k,2k}=0.
\end{eqnarray}
\textbf{2}. \qquad $\eta_{_{2i}}=\delta_{ik}$ , $\alpha=0 ,
\beta=1$\quad\quad
\begin{eqnarray}\label{48}
|\vartheta\rangle=|2k,2k+1\rangle \longrightarrow
\langle\vartheta|\varphi\rangle=a_{2k,2k+1}=0.
\end{eqnarray}
\textbf{3}. \qquad $\eta_{_{2i+1}}=\delta_{ik}$ , $\alpha=0 ,
\beta=1$\quad\quad
\begin{eqnarray}\label{49}
|\vartheta\rangle=|2k+1,2k\rangle \longrightarrow
\langle\vartheta|\varphi\rangle=a_{2k+1,2k}=0.
\end{eqnarray} \textbf{4}. \qquad $\eta_{_{2i+1}}=\delta_{ik}$ , $\alpha=1 ,
\beta=0$\quad\quad
\begin{eqnarray}\label{50}
|\vartheta\rangle=|2k+1,2k+1\rangle \longrightarrow
\langle\vartheta|\varphi\rangle=a_{2k+1,2k+1}=0.
\end{eqnarray}\textbf{5}. \qquad
$\eta_{_{2k}}=\alpha' , \eta_{_{2l}}=\beta'$ ,
$|\alpha'|^{2}+|\beta'|^{2}=1$ , $k<l$ , $\alpha=0 , \beta=1$
$$|\vartheta\rangle=|\alpha'|^{2}|2k,2k+1\rangle+\alpha'\beta'^{\ast}|2l,2k+1\rangle+\beta'\alpha'^{\ast}|2k,2l+1\rangle+|\beta'|^{2}|2l,2l+1\rangle,$$
$$\langle\vartheta|\varphi\rangle=\alpha'\beta'^{\ast}a_{2l,2k+1}+\beta'\alpha'^{\ast}a_{2k,2l+1}=0$$
\be\label{51} \longrightarrow a_{2l,2k+1}=0\quad,\quad
a_{2k,2l+1}=0.\ee\textbf{6}. \qquad $\eta_{_{2k+1}}=\alpha' ,
\eta_{_{2l+1}}=\beta'$ , $|\alpha'|^{2}+|\beta'|^{2}=1$ , $k<l$ ,
$\alpha=0 , \beta=1$
$$|\vartheta\rangle=|\alpha'|^{2}|2k+1,2k\rangle+\alpha'\beta'^{\ast}|2l+1,2k\rangle+\beta'\alpha'^{\ast}|2k+1,2l\rangle+|\beta'|^{2}|2l+1,2l\rangle,$$
$$\langle\vartheta|\varphi\rangle=\alpha'\beta'^{\ast}a_{2l+1,2k}+\beta'\alpha'^{\ast}a_{2k+1,2l}=0$$
\be\label{52} \longrightarrow a_{2l+1,2k}=0,\quad\quad
a_{2k+1,2l}=0.\ee \textbf{7}. \qquad $\eta_{_{2k}}=\alpha' ,
\eta_{_{2l+1}}=\beta'$ , $|\alpha'|^{2}+|\beta'|^{2}=1$ , $k<l$ ,
$\alpha=0 , \beta=1$
$$|\vartheta\rangle=-|\alpha'|^{2}|2k,2k+1\rangle-\alpha'\beta'^{\ast}|2l+1,2k+1\rangle+\beta'\alpha'^{\ast}|2k,2l\rangle+|\beta'|^{2}|2l+1,2l\rangle,$$
$$\langle\vartheta|\varphi\rangle=-\alpha'\beta'^{\ast}a_{2l+1,2k}+\beta'\alpha'^{\ast}a_{2k,2l}=0$$
\be\label{53} \longrightarrow a_{2l+1,2k+1}=0.\quad\quad
a_{2k,2l}=0.\ee \textbf{8}. \qquad $\eta_{_{2k+1}}=\alpha' ,
\eta_{_{2l}}=\beta'$ , $|\alpha'|^{2}+|\beta'|^{2}=1$ , $k<l$ ,
$\alpha=0 , \beta=1$
$$|\vartheta\rangle=|\alpha'|^{2}|2k+1,2k\rangle+\alpha'\beta'^{\ast}|2l,2k\rangle-\beta'\alpha'^{\ast}|2k+1,2l+1\rangle-|\beta'|^{2}|2l,2l+1\rangle,$$
$$\langle\vartheta|\varphi\rangle=\alpha'\beta'^{\ast}a_{2l,2k}-\beta'\alpha'^{\ast}a_{2k+1,2l+1}=0$$
\be\label{54} \longrightarrow a_{2k+1,2l+1}=0,\quad\quad
a_{2l,2k}=0.\ee therefore, $\mathrm{Q}=0$ on the $2n\otimes2n$ subspace. From the other side, since all of the separable states which lie on the complement subspace of the $2n\otimes2n$ one belong to the $\mathrm{P}_{\mathrm{W}_{OPC}}$ (see appendix B), then it is obvious that Q is also zero on that subspace. Consequently $\mathrm{Q}=0$ on $d\otimes d$ Hilbert space hence $\mathrm{W}_{OPC}$ is optimal nd-EW. Clearly, if J is full-rank,
then $\mathrm{O}_{_{1}}^{T_{A}}$ and $\mathrm{O}_{_{2}}^{T_{A}}$
will be zero so $\mathrm{W}_{C}=\mathrm{W}_{OPC}$ and
$\mathrm{W}_{OPC}$ is an optimal nd-EW on the $d\otimes d$ Hilbert
space. It is concluded that when J is full-rank, $\mathrm{W}_{C}$
is optimal nd-EW and when J is not full-rank $\mathrm{W}_{C}$ is
not optimal nd-EW; therefore, despite the optimality, the
nd-optimality of canonical EWs depends on the rank of J.

\subsection{The Positive Maps Corresponding to Canonical EWs}
Since the nd-EWs have an essential role in the studying of separability problem in quantum theory, by using Jamiolkowski isomorphism
\cite{jamiol} between operators and maps, the non-decomposable
positive maps (or nd-positive maps) have the same role as nd-EWs.
By this isomorphism, one can obtain the corresponding positive map
of the canonical EW $\mathrm{W}_{C}\in \mathcal{H}_{d}\otimes
\mathcal{H}_{d}$ (\ref{eq12}) as discussed in subsection (2.1).
Consider the following equation
\begin{eqnarray}\label{55}
\phi(\rho)=\mathrm{Tr}_{\mathrm{B}}(\mathrm{W}_{C}(\mathrm{I_{d}}\otimes
\rho^{T})).
\end{eqnarray}
Where $\rho$ is a density operator on the d-dimensional Hilbert
space. This equation shows how to construct the map $\phi$ from a
given operator $\mathrm{W}_{C}$. After some
calculations, the following result is obtained
\begin{eqnarray}\label{56}
\phi(\rho)=\mathrm{I_{d}}\mathrm{Tr}(\rho)-\rho-\mathrm{J}^{T}\rho^{T}\mathrm{J}.
\end{eqnarray}From the properties of $\mathrm{W}_{C}$ discussed earlier, we
expect that, if the $rank(J)<4$, then the $\phi(\rho)$ is
decomposable positive map (or d-positive map), especially when
$rank(J)=0$, $\phi(\rho)$ is the well-known reduction map
\cite{M.P}. Therefore, for $rank(J)\geq4$, $\phi(\rho)$ is
nd-positive map.

\section{Entanglement Witnesses Corresponding to possible partitions of J}

In this section, by referring to each possible partition of J, we are going to construct a new set of canonical EWs.
Therefore, for a given J, we have a set of canonical EWs
corresponding to the set of possible partitions of J. It is shown
that, for a given partition of J, a PPT state is constructed for
that partition. The entanglement of this PPT state in some range of
parameters is detected by the corresponding canonical EW
established in the same partition. Suppose that J is full-rank, i.e., $d=2n$. Consider a partition of d-dimensional single party
Hilbert space $\mathcal{H}$ to its $2\mu_{i}$-dimensional
subspaces, $\mathcal{H}_{2\mu_{i}}$s, through the following direct
sum
\begin{eqnarray}\label{eq57}
\mathcal{H}=\mathcal{H}_{2\mu_{1}}\oplus
\mathcal{H}_{2\mu_{2}}\oplus
\mathcal{H}_{2\mu_{3}}\oplus\cdots\oplus \mathcal{H}_{2\mu_{\nu}}.
\end{eqnarray}
Also consider the following Hermitian operator
$$\mathrm{W_{C}}=\mathrm{I}_{d}\otimes\mathrm{I_{d}}-d|\psi\rangle\langle\psi|-d(\mathrm{U}_{1}^{T}\otimes
\mathrm{I_{d}})|\psi\rangle\langle\psi|^{T_{A}}(\mathrm{U}_{1}\otimes
\mathrm{I_{d}})-d(\mathrm{U}_{2}^{T}\otimes
\mathrm{I_{d}})|\psi\rangle\langle\psi|^{T_{A}}(\mathrm{U}_{2}\otimes
\mathrm{I}_{d})$$ \be\label{eq58}-d(\mathrm{U}_{3}^{T}\otimes
\mathrm{I_{d}})|\psi\rangle\langle\psi|^{T_{A}}(\mathrm{U}_{3}\otimes
\mathrm{I}_{d})-\cdots-d(\mathrm{U}_{\nu}^{T}\otimes
\mathrm{I}_{d})|\psi\rangle\langle\psi|^{T_{A}}(\mathrm{U}_{\nu}\otimes
\mathrm{I}_{d}),\ee in which every $\mathrm{U}_{i}$ is
block-diagonal full-rank matrix on $\mathcal{H}_{2\mu_{i}}$
($k=1,\cdots,\nu$) such that
\begin{eqnarray}\label{eq59}
\mathrm{J}=\mathrm{U}_{1}+\mathrm{U}_{2}+\mathrm{U}_{3}+\cdots+\mathrm{U}_{\nu},
\end{eqnarray}
where $\mathrm{U}_{i}\mathrm{U}_{j}=0$ ($i\neq j=1,...,\nu$). Each $\mathrm{U}_{i}$ is a full-rank canonical skew-symmetric matrix in $\mathcal{H}_{2\mu_{i}}$. Therefore the rank of $\mathrm{J}$ is the sum of
the ranks of $\mathrm{U}_{i}$s that is
\begin{eqnarray}\label{eq60}
2n=2\mu_{1}+2\mu_{2}+2\mu_{3}+\cdots+2\mu_{\nu}
\end{eqnarray}
hence we obtain the next result
\begin{eqnarray}\label{eq61}
n=\mu_{1}+\mu_{2}+\mu_{3}+\cdots+\mu_{\nu}.
\end{eqnarray}
It is well-known that the numbers
($\mu_{1},\mu_{2},\mu_{3},\cdots,\mu_{\nu}$) are a partition of
$n$. Generally from \cite{nathanson}, for a given number $n$ there
are p(n) number of partitions
($\mu_{1},\mu_{2},\mu_{3},\cdots,\mu_{\nu}$) with
$\mu_{1}\geq\mu_{2}\geq\mu_{3}\geq,\cdots,\geq\mu_{\nu}$ (for
example consider 5, the number of partitions for it, is $p(5)=7$).
Hence we say that the $\mathrm{U}_{k}$s in (\ref{eq59}) are a
partition of $\mathrm{J}$. Therefore for the other possible
partitions of $n$, we have corresponding partitions for
$\mathrm{J}$ and corresponding Hermitian operators such as
$\mathrm{W_{C}}$ which is renamed as
$\mathrm{W_{C}}(\mu_{1},\mu_{2},\mu_{3},\cdots,\mu_{\nu})$.
\par In the same way, as described in section 2, the
expectation values of Hermitian operator $\mathrm{W_{C}}$ with all
product states are given as
$$\langle\eta|\otimes\langle\zeta|\mathrm{W_{C}}(\mu_{1},\mu_{2},\cdots,\mu_{\nu-1},\mu_{\nu})|\eta\rangle\otimes|\zeta\rangle=1-|\langle\zeta|\eta^{\ast}\rangle|^{2}-|\langle\zeta|\mathrm{U}_{1}|\eta\rangle|^{2}
-|\langle\zeta|\mathrm{U}_{2}|\eta\rangle|^{2}$$
\be\label{62}-|\langle\zeta|\mathrm{U}_{3}|\eta\rangle|^{2}-\cdots-|\langle\zeta|\mathrm{U}_{\nu}|\eta\rangle|^{2},
\ee in which the states \{$|\eta^{\ast}\rangle$ ,
$\mathrm{U}_{1}|\eta\rangle$ , $\mathrm{U}_{2}|\eta\rangle$ ,
$\mathrm{U}_{3}|\eta\rangle$, $\cdots$ ,
$\mathrm{U}_{\nu}|\eta\rangle$\} are orthogonal to each other.
If $\mathrm{U}_{i}|\eta\rangle$, for each i, is normalized then $|\zeta\rangle$ can be written as $|\zeta\rangle=\alpha|\eta^{\ast}\rangle+\sum_{i=0}^{\nu}\beta_{i} U_{i}|\eta\rangle$ with $|\alpha|^{2}+\sum_{i=1}^{\nu}|\beta_{i}|^{2}=1$ otherwise it can not.
Therefore $\mathrm{W_{C}}(\mu_{1},\mu_{2},\mu_{3},\cdots,\mu_{\nu})$ has
positive expectation values with all separable states so it can be
considered as a canonical EW for the partition
($\mu_{1},\mu_{2},\mu_{3},\cdots,\mu_{\nu}$).
par The action of $\mathrm{U}_{i}$ ($i=1,...,\nu$) on the basis states of $\mathrm{H}_{2\mu_{i}}$ ($i=1,...,\nu$) is as
$$\mathrm{U}_{i}|2k\rangle=-|2k+1\rangle, \qquad\
\mathrm{U}_{i}|2k+1\rangle=|2k\rangle,$$
\be\label{63}i=1,...,\nu, \quad\quad k=0,...,\mu_{i}-1,\ee
therefore $\mathrm{W}_{C}(\mu_{1},\mu_{2},\mu_{3},\cdots,\mu_{\nu})$ in equation (\ref{eq58}) for a given partition ($\mu_{1},\mu_{2},\mu_{3},\cdots,\mu_{\nu}$)
can be obtained as
$$\mathrm{W_{C}}(\mu_{1},\mu_{2},\mu_{3},\cdots,\mu_{\nu})=\mathcal{W}_{C}(\mu_{1},\mu_{2},\mu_{3},\cdots,\mu_{\nu})$$
\be\label{eq64}+\mathrm{O}^{t_{A}}(\mu_{1})+\mathrm{O}^{t_{A}}(\mu_{2})+\mathrm{O}^{t_{A}}(\mu_{3})+\cdots+\mathrm{O}^{t_{A}}(\mu_{\nu-1}),\ee
where
\begin{eqnarray}\label{65}
\mathcal{W}_{C}(\mu_{1},\mu_{2},\mu_{3},\cdots,\mu_{\nu})=\mathrm{W_{C}}(\mu_{1})\oplus
\mathrm{W_{C}}(\mu_{2})\oplus \mathrm{W_{C}}(\mu_{3})\oplus\cdots
\oplus \mathrm{W_{C}}(\mu_{\nu})
\end{eqnarray}
The operators $\mathrm{W_{C}}(\mu_{i})$ ($i=1,\cdots,\nu$) and
$\mathrm{O}^{t_{A}}(\mu_{i})$ ($i=1,\cdots,\nu-1$) have been given
in the appendix C. By the same prescription shown in section (2.3),
the $\mathrm{W}_{C}(\mu_{1},\mu_{2},\mu_{3},\cdots,\mu_{\nu})$ is
optimal EW for each partition of n. Each $\mathrm{W_{C}}(\mu_{i})$
($i=1,\cdots,\nu$) is an EW in the $2\mu_{i}\otimes2\mu_{i}$
subspace of the $d\otimes d$ Hilbert space. Also for every $\mu_{i}>1$ the corresponding
$\mathrm{W}_{C}(\mu_{i})$ is an optimal nd-EW and hence
$\mathcal{W}_{c}(\mu_{1},\mu_{2},\mu_{3},\cdots,\mu_{\nu})$ which
is the direct some of the $\mathrm{W_{C}}(\mu _{i})$s, is also an
optimal nd-EW. The operators $\mathrm{O}(\mu_{i})$
$(i=1,...,\nu-1)$ are positive operators then the operators
$\mathrm{O}^{t_{A}}(\mu_{i})$ ($i=1,\cdots,\nu-1$) are optimal
d-EWs.
\par As mentioned in section (2.1), if a given $\mu_{i}$ be equal to one then the
corresponding $\mathrm{W}_{C}(\mu _{i})$ will be zero. Therefore
if all of $\mu_{i}$s become one, i.e. for the partition
$(1,1,1,\cdots,1)$, then $\mathrm{W}_{C}(1,1,1,\cdots,1)$ will be
an optimal d-EW. The witness corresponding to the partition
($n=\mu_{1}$), i.e. $\mathrm{W}_{C}(\mu_{1})$, is an optimal
nd-EW which was discussed earlier. In the end, the witnesses which
correspond to the other partitions between these two partitions
, as the equation (\ref{eq64}), are a mixture of optimal d-EWs and the optimal nd-EWs.
Therefore, by considering the optimality of the nd-EWs discussed in
subsection (2.3), these witnesses are not optimal nd-EWs.
\par In the next step, for a given partition ($\mu_{1},\mu_{2},\mu_{3},\cdots\mu_{\nu}$),
a $\mathrm{PPT}$ state is introduced. This state can be entangled
and the signature of entanglement for it is shown by the witness
which corresponds to the same partition discussed above. This
state is the following one
$$\rho(\mu_{1},\mu_{2},\mu_{3},\cdots,\mu_{\nu})$$
\be\label{66}=\frac{1}{\mathcal{N}(\mu_{1},\mu_{2},\mu_{3},\cdots,\mu_{\nu})}(\varrho(\mu_{1},\mu_{2},\mu_{3},\cdots,\mu_{\nu})+\sigma(\mu_{1})+\sigma(\mu_{1})+\sigma(\mu_{3})+\cdots+\sigma(\mu_{\nu-1}))\ee
such that
\begin{eqnarray}\label{67}
\varrho(\mu_{1},\mu_{2},\mu_{3},\cdots,\mu_{\nu})=\rho(\mu_{1})\oplus
\rho(\mu_{2})\oplus\rho(\mu_{3})\oplus\cdots \oplus
\rho(\mu_{\nu}),
\end{eqnarray}
where $\mathcal{N}(\mu_{1},\mu_{2},\mu_{3},\cdots,\mu_{\nu})$ is
the norm of the $\rho(\mu_{1},\mu_{2},\mu_{3},\cdots,\mu_{\nu})$.
The operators $\rho(\mu_{i})$s ($i=1,\cdots,\nu$) are also
(unnormalized) $\mathrm{PPT}$ states in the subspace
$2\mu_{i}\otimes2\mu_{i}$ and by the witnesses $\mathrm{W}_{C}(\mu_{i})$s, they are
entangled PPT states. Also the operator
$\varrho(\mu_{1},\mu_{2},\mu_{3},\cdots,\mu_{\nu})$ which is the
direct sum of the $\rho(\mu_{i})$s, is $\mathrm{PPT}$ state so its
entanglement is detected by the witness
$\mathcal{W}_{c}(\mu_{1},\mu_{2},\mu_{3},\cdots,\mu_{\nu})$ and
finally $\sigma(\mu_{i})$s ($i=1,\cdots,\nu-1$) are Hermitian
operators. All these operators together with the positivity and
$\mathrm{PPT}$ conditions for
$\rho(\mu_{1},\mu_{2},\mu_{3},\cdots,\mu_{\nu})$ have been given
in appendix D.
\par Now at the end of this section the expectation value of $\mathrm{W}_{C}(\mu_{1},\mu_{2},\mu_{3},\cdots,\mu_{\nu})$
with respect to the $\mathrm{PPT}$ state
$\rho(\mu_{1},\mu_{2},\mu_{3},\cdots,\mu_{\nu})$, by considering
both positivity and $\mathrm{PPT}$ conditions, is calculated. In
the same way as in subsection (2.2), after some calculations we
obtain the following lower bound
$$\mathrm{Tr}[\mathrm{W}_{C}(\mu_{1},\mu_{2},\mu_{3},\cdots,\mu_{\nu})\rho(\mu_{1},\mu_{2},\mu_{3},\cdots,\mu_{\nu})]$$
\be\label{68}\geq-\frac{d}{\sum_{\epsilon=1}^{\nu}(2\mu_{\epsilon}(2\mu_{\epsilon}+1)+4\mu_{\epsilon}(d-2\sum_{\theta=1}^{\epsilon}\mu_{_{\theta}}))}.\ee
Therefore we say that if a $\mathrm{PPT}$ state
$\rho(\mu_{1},\mu_{2},\mu_{3},\cdots,\mu_{\nu})$ satisfy the
following inequality
$$-\frac{d}{\sum_{\epsilon=1}^{\nu}(2\mu_{\epsilon}(2\mu_{\epsilon}+1)+4\mu_{\epsilon}(d-2\sum_{\theta=1}^{\epsilon}\mu_{_{\theta}}))}$$
\be\label{69}\leq
  \mathrm{Tr}[\mathrm{W}_{C}(\mu_{1},\mu_{2},\mu_{3},\cdots,\mu_{\nu})\rho(\mu_{1},\mu_{2},\mu_{3},\cdots,\mu_{\nu})]<0
,\ee then it is an entangled $\mathrm{PPT}$ state detected by the
witness
$\mathrm{W}_{C}(\mu_{1},\mu_{2},\mu_{3},\cdots,\mu_{\nu})$.
Finally if $\rho(\mu_{1},\mu_{2},\mu_{3},\cdots,\mu_{\nu})$
violate the $\mathrm{PPT}$ condition, it always satisfies the next
inequality
$$-\frac{\sum_{\epsilon=1}^{\nu}(2\mu_{\epsilon}(\mu_{\epsilon}-1)+2\mu_{\epsilon}(d-2\sum_{\theta=1}^{\epsilon}\mu_{_{\theta}}))}{d}\leq\mathrm{Tr}[\mathrm{W}_{C}(\mu_{1},\mu_{2},\mu_{3},\cdots,\mu_{\nu})\rho(\mu_{1},\mu_{2},\mu_{3},\cdots,\mu_{\nu})]$$
\be\label{70}<-\frac{d}{\sum_{\epsilon=1}^{\nu}(2\mu_{\epsilon}(2\mu_{\epsilon}+1)+4\mu_{\epsilon}(d-2\sum_{\theta=1}^{\epsilon}\mu_{_{\theta}}))}.\ee

\section{Extensions}
In this section, at first, we extend our approach used for $d\otimes d$ cases in
the previous sections to obtain EWs in $d_{1}\otimes d_{2}$
quantum systems ($d_{1}<d_{2}$). Secondly, we extend the idea of canonical EW introduced in subsection (2.1) to construct other EWs. Let us
introduce a projection operator $\mathrm{P}_{c}$ on the
$d_{2}$-dimensional party in which $\mathrm{Im}(\mathrm{P}_{c})$ is a $d_{1}$-dimensional Hilbert space. Since in the
$d_{2}$-dimensional vector space one can construct
$d_{1}$-dimensional subspace by $\mathrm{C}_{d_{1}}^{d_{2}}$ (the number of combinations of $d_{2}$
distinct objects taken $d_{1}$ at a time without repetitions)
distinct ways; therefore, the subscript $c$
($c=1,\cdots,\mathrm{C}_{d_{1}}^{d_{2}}$) dentes a projection
operator among the set of $\mathrm{C}_{d_{1}}^{d_{2}}$ ones . From
the linear algebra, we know that
$\mathcal{H}_{d_{2}}=\mathrm{Ker}\mathrm{P}_{c}\oplus
\mathrm{Im}\mathrm{P}_{c}$.
 Therefore for
every $|\alpha\rangle\in \mathcal{H}_{d_{2}}$, we have
$|\alpha\rangle=|\alpha_{c}\rangle+|\alpha_{c^{'}}\rangle$ where
$|\alpha_{c}\rangle\in \mathrm{Im}\mathrm{P}_{c}$ and
$|\alpha_{c^{'}}\rangle\in \mathrm{Ker}\mathrm{P}_{c}$ with
dim($\mathrm{Im}\mathrm{P}_{c}$)=$d_{1}$ and
dim($\mathrm{Ker}\mathrm{P}_{c}$)=$d_{2}-d_{1}$. Now we introduce
the following Hermitian operator
\begin{equation}\label{71}
\mathrm{W}=\mathrm{I}_{d_{1}}\otimes
\mathrm{I}_{c}-d_{1}|\psi_{c}\rangle\langle\psi_{c}|-d_{1}(\mathrm{U}_{d_{1}}^{T}\otimes
\mathrm{I}_{c})|\psi_{c}\rangle\langle\psi_{c}|^{T_{A}}(\mathrm{U}_{d_{1}}\otimes
\mathrm{I}_{c})
\end{equation}
where
$\mathrm{I}_{c}=\sum_{_{i=0}}^{^{d_{1}-1}}|i_{c}\rangle\langle
i_{c}|$ is the identity operator in the projected subspace of
$\mathrm{H}_{d_{2}}$ corresponding to a combination denoted by
$c$. So the identity operator in the corresponding $d_{1}\otimes
d_{1}$ subspace of $d_{1}\otimes d_{2}$ system is given as
\begin{equation}\label{72}
\mathrm{I}_{d_{1}}\otimes
\mathrm{I}_{c}=\sum_{i,j=0}^{d_{1}-1}|i,j_{c}\rangle\langle
i,j_{c}|
\end{equation} and $|\psi_{c}\rangle$
is the the maximal entangled Bell-state in that subspace
\begin{equation}\label{73}
|\psi_{c}\rangle=\frac{1}{\sqrt{d_{1}}}\sum_{i=0}^{d_{1}-1}|i,i_{c}\rangle,
\end{equation}
in which $|i_{c}\rangle\in \mathrm{Im}\mathrm{P}_{c}$ for
$i=0,\cdots,d_{1}-1$. It should
be noted that the skew-symmetric operator $\mathrm{U}_{d_{1}}$ is
defined on the $d_{1}$-dimensional party of the $d_{1}\otimes
d_{2}$ system. To clarify, we give an example in
$4\otimes5$. Sine $C_{4}^{5}=5$ then we have five $4$-dimensional
projected subspace for $\mathrm{H}_{5}$. The first one spanned by
$\{|0_{1}\rangle=|0\rangle, |1_{1}\rangle=|1\rangle,
|2_{1}\rangle=|2\rangle, |3_{1}\rangle=|3\rangle\}$, the second
one by $\{|0_{2}\rangle=|0\rangle, |1_{2}\rangle=|1\rangle,
|2_{2}\rangle=|2\rangle, |3_{2}\rangle=|4\rangle\}$, the third one
by $\{|0_{3}\rangle=|0\rangle, |1_{3}\rangle=|2\rangle,
|2_{3}\rangle=|3\rangle, |3_{3}\rangle=|4\rangle\}$, the fourth
one by $\{|0_{4}\rangle=|0\rangle, |1_{4}\rangle=|1\rangle,
|2_{4}\rangle=|3\rangle, |3_{4}\rangle=|4\rangle\}$ and the fifth
one by $\{|0_{5}\rangle=|1\rangle, |1_{5}\rangle=|2\rangle,
|2_{5}\rangle=|3\rangle, |3_{5}\rangle=|4\rangle\}$. Therefore one
has five ways to construct the operator (\ref{71}). It is clear that the operator (\ref{71}) is the
same operator (\ref{eq4}) in $d_{1}\otimes d_{1}$ bipartite
quantum system which has been embedded in $d_{1}\otimes d_{2}$
Hilbert space. Consequently, the number of such operators obtained
by embedding in this way is, in fact, $C_{d_{1}}^{d_{2}}$.
\par By the same arguments proposed in section 2, the
operator (\ref{71}) can be considered as an entanglement witness
which is defined on the $d_{1}\otimes d_{2}$ Hilbert space. Using
$\mathrm{J}_{d_{1}}$ instead of $\mathrm{U}_{_{d_{1}}}$ in
(\ref{71}), we obtain a canonical form for W as
\begin{equation}\label{eq74}
\mathrm{W}_{C}=\mathrm{I}_{d_{1}}\otimes
\mathrm{I}_{c}-d_{1}|\psi_{c}\rangle\langle\psi_{c}|-d_{1}(\mathrm{J^{T}_{d_{1}}}\otimes
\mathrm{I}_{c})|\psi_{c}\rangle\langle\psi_{c}|^{T_{A}}(\mathrm{J_{d_{1}}}\otimes
\mathrm{I}_{c}),
\end{equation}
where $\mathrm{Q_{d_{1}}}$ is an orthogonal matrix
($\mathrm{Q_{d_{1}}}\mathrm{Q^{T}_{_{d_{1}}}}=\mathrm{Q^{T}_{_{d_{1}}}}\mathrm{Q_{{d_{1}}}}=\mathrm{I}_{_{d_{1}}}$)
and $\mathrm{J}_{_{d_{1}}}$ is the canonical form of
$\mathrm{U}_{_{d_{1}}}$ i.e.,
$\mathrm{U}_{_{d_{1}}}=\mathrm{Q}_{_{d_{1}}}\mathrm{J}_{_{d_{1}}}\mathrm{Q}^{T}_{_{d_{1}}}$.
The action of $\mathrm{J}_{d_{1}}$ (whose rank is $2n_{1}$) on
the basis of the $d_{1}$-dimensional single party Hilbert space is
similar to (\ref{eq17}) and (\ref{eq18}). Consequently, after some
calculations, the expanded form of $\mathrm{W}_{C}$ becomes
$$\mathrm{W}_{C}=\sum_{_{i\neq j=0}}^{n_{1}-1}(|2i,(2j)_{_{c}}\rangle\langle2i,(2j)_{_{c}}|+|2i,(2j+1)_{_{c}}\rangle\langle2i,(2j+1)_{_{c}}|+|2i+1,(2j)_{_{c}}\rangle\langle2i+1,(2j)_{_{c}}|$$
$$+|2i+1,(2j+1)_{_{c}}\rangle\langle2i+1,(2j+1)_{_{c}}|-|2i,(2i)_{_{c}}\rangle\langle2j,(2j)_{_{c}}|-|2i,(2i)_{_{c}}\rangle\langle2j+1,(2j+1)_{_{c}}|$$
$$-|2i+1,(2i+1)_{_{c}}\rangle\langle2j,(2j)_{_{c}}|-|2i+1,(2i+1)_{_{c}}\rangle\langle2j+1,(2j+1)_{_{c}}|$$
$$-|2i+1,(2j)_{_{c}}\rangle\langle2j+1,(2i)_{_{c}}|+|2i+1,(2j+1)_{_{c}}\rangle\langle2j,(2i)_{_{c}}|$$
$$+|2i,(2j)_{_{c}}\rangle\langle2j+1,(2i+1)_{_{c}}|-|2i,(2j+1)_{_{c}}\rangle\langle2j,(2i+1)_{_{c}}|)$$
$$+\sum_{_{i=0}}^{n_{1}-1}\sum_{_{j=2n_{1}}}^{d_{1}-1}(|2i,j_{_{c}}\rangle\langle2i,j_{_{c}}|+|2i+1,j_{_{c}}\rangle\langle2i+1,j_{_{c}}|+|j,(2i)_{_{c}}\rangle\langle
j,(2i)_{_{c}}|+|j,(2i+1)_{_{c}}\rangle\langle j,(2i+1)_{_{c}}|$$
$$-|2i,(2i)_{_{c}}\rangle\langle j,j_{_{c}}|-|2i+1,(2i+1)_{_{c}}\rangle\langle
j,j_{_{c}}|-|j,j_{_{c}}\rangle\langle
2i,(2i)_{_{c}}|-|j,j_{_{c}}\rangle\langle 2i+1,(2i+1)_{_{c}}|)$$
\be\label{75}+\sum_{_{i,j=2n}}^{d_{1}-1}(|i,j_{_{c}}\rangle\langle
i,j_{_{c}}|-|i,i_{_{c}}\rangle\langle j,j_{_{c}}|).\ee
These witnesses are optimal d-$\mathrm{EW}$s for $d_{1}<4$; otherwise, they are nd-$\mathrm{EW}$s, so they can detect
the entanglement of some $\mathrm{PPT}$ states. Optimality and
nd-optimality problem for these witnesses is similar to those ones
which were discussed in subsection (2.3). Now we
introduce $\mathrm{PPT}$ states in which some of them are entangled.
Consider the following class of states
$$\rho=\frac{1}{\mathcal{N}}(a_{0}|\psi_{c}\rangle\langle\psi_{c}|+a_{0}\sum_{i=0}^{n_{1}-1}(|2i,(2i)_{_{c}}\rangle\langle2i,(2i)_{_{c}}|+|2i+1,(2i+1)_{_{c}}\rangle\langle2i+1,(2i+1)_{_{c}}|$$
$$-|2i,(2i)_{_{c}}\rangle\langle2i+1,(2i+1)_{_{c}}|-|2i+1,(2i+1)_{_{c}}\rangle\langle2i,(2i)_{_{c}}|)$$
$$+\sum_{i=0}^{n_{1}-1}(a_{2i+2,(2i)_{_{c}}}|2i+2,(2i)_{_{c}}\rangle\langle2i+2,(2i)_{_{c}}|+a_{2i+1,(2i+3)_{_{c}}}|2i+1,(2i+3)_{_{c}}\rangle\langle2i+1,(2i+3)_{_{c}}|$$
$$-C_{i}(|2i+2,(2i)_{_{c}}\rangle\langle2i+1,(2i+3)_{_{c}}|+|2i+1,(2i+3)_{_{c}}\rangle\langle2i+2,(2i)_{_{c}}|))$$
$$+\sum_{i\neq j=0,i-j\neq 1,j-i\neq
n_{1}-1}^{n_{1}-1}a_{2i,(2j)_{_{c}}}|2i,(2j)_{_{c}}\rangle\langle2i,(2j)_{_{c}}|$$
$$+\sum_{i\neq j=0,j-i\neq 1,i-j\neq n_{1}-1}^{n_{1}-1}a_{2i+1,(2j+1)_{_{c}}}|2i+1,(2j+1)_{_{c}}\rangle\langle2i+1,(2j+1)_{_{c}}|$$
$$+\sum_{i,j=0}^{n_{1}-1}(a_{2i,(2j+1)_{_{c}}}|2i,(2j+1)_{_{c}}\rangle\langle2i,(2j+1)_{_{c}}|+a_{2i+1,(2j)_{_{c}}}|2i+1,(2j)_{_{c}}\rangle\langle2i+1,(2j)_{_{c}}|)$$
$$+\sum_{i=2n_{1}}^{d_{1}-1}\sum_{j=0}^{n_{1}-1}a_{2j,i_{_{c}}}(|2j,i_{_{c}}\rangle\langle2j,i_{_{c}}|+a_{2j+1,i_{_{c}}}|2j+1,i_{_{c}}\rangle\langle2j+1,i_{_{c}}|$$
$$+a_{i,(2j)_{_{c}}}|i,(2j)_{_{c}}\rangle\langle i,(2j)_{_{c}}|+a_{i,(2j+1)_{_{c}}}|i,(2j+1)_{_{c}}\rangle\langle
i,(2j+1)_{_{c}}|)$$
$$+\sum_{i\neq j=2n_{1}}^{d_{1}-1}a_{i,j_{_{c}}}|i,j_{_{c}}\rangle\langle i,j_{_{c}}|$$
\be\label{76}+\sum_{i=0}^{d_{1}-1}\sum_{j=0}^{d_{2}-d_{1}-1}a_{i,j_{c^{'}}}|i,j_{c^{'}}\rangle\langle
i,j_{c^{'}}|,\ee it is clear that the number of such states is
$C^{d_{2}}_{d_{1}}$. The positivity and $\mathrm{PPT}$ conditions
are the same as for $d\otimes d$ cases except that we have an
additional condition for positivity through the inequalities
$a_{i,j_{_{c^{'}}}}\geq0$ ($i=0,\cdots,d_{1}-1$ ,
$j=0,\cdots,d_{2}-d_{1}-1$). The subscript $c^{'}$ denotes those
states which lie in the $\mathrm{Ker}\mathrm{P}_{c}$
($i=0,\cdots,d_{2}-d_{1}-1$). Now the expectation value of the
canonical $\mathrm{EW}$ with respect to the $\mathrm{PPT}$ state
$\rho$ gives out the following lower bound
\begin{eqnarray}\label{eq77}
\mathrm{Tr}(\mathrm{W}_{C}\rho)\geq\frac{-2n_{1}}{d_{1}+4n_{1}^{2}+(d_{1}-2n_{1})(d_{1}+2n_{1}-1)},\qquad
 \qquad n_{1}=2,3,4,\cdots.
\end{eqnarray}
Therefore if those PPT states satisfy the following inequality
\begin{eqnarray}\label{eq78}
\frac{-2n_{1}}{d_{1}+4n^{2}_{1}+(d_{1}-2n_{1})(d_{1}+2n_{1}-1)}\leq\mathrm{Tr}(\mathrm{W}_{C}\rho)<0,\qquad
 \qquad n_{1}=2,3,4,\cdots,
\end{eqnarray}
then they are entangled. And, as before, if the states $\rho$
violate the PPT conditions, then they always fulfill the following
inequality
$$-\frac{4n_{1}(n_{1}-1)+(d_{1}-2n_{1})(d_{1}+2n_{1}-1)}{d_{1}+2n_{1}}\leq\mathrm{Tr}(\mathrm{W}_{C}\rho)<\frac{-2n_{1}}{d_{1}+4n^{2}_{1}+(d_{1}-2n_{1})(d_{1}+2n_{1}-1)},$$
\be\label{79}n_{1}=2,3,4,\cdots.\ee In the end of this paper, the idea of canonical EW introduced in equation (\ref{eq12}) can be extended to construct other EWs. Let us , at first, assume that the single party Hilbert spaces  $\mathcal{H}_{1}$ and $\mathcal{H}_{2}$ and the tensor product Hilbert space $\mathcal{H}_{1}\otimes\mathcal{H}_{2}$ are defined on real field. As an illustration to this restriction, any entangled state which lie in a real tensor product Hilbert space $\mathcal{H}_{1}\otimes\mathcal{H}_{2}$ can be generated, by interactions or any entanglement generating process, from single party states which lie in real single party Hilbert spaces $\mathcal{H}_{1}$ and $\mathcal{H}_{2}$. It should be noted that any LOCC (local operation and classical communication) must be restricted on the real field. Therefore by these considerations, when we deal to detect the entanglement of a state which lies in a real tensor product Hilbert space $\mathcal{H}_{1}\otimes\mathcal{H}_{2}$ by constructing an EW, it is sufficient that our EW should have positive expectation value with respect to all real separable states. Now consider the following operator
$$\mathrm{W}=\mathrm{I}_{d}\otimes
\mathrm{I}_{d}-d|\psi\rangle\langle\psi|-d(\mathrm{J}^{T}\otimes
\mathrm{I}_{d})|\psi\rangle\langle\psi|^{T_{A}}(\mathrm{J}\otimes
\mathrm{I}_{d})$$
\be\label{eq80}-d(\mathrm{J^{'}}^{T}\otimes
\mathrm{I}_{d})|\psi\rangle\langle\psi|^{T_{A}}(\mathrm{J^{'}}\otimes
\mathrm{I}_{d})-d(\mathrm{J^{''}}^{T}\otimes
\mathrm{I}_{d})|\psi\rangle\langle\psi|^{T_{A}}(\mathrm{J^{''}}\otimes
\mathrm{I}_{d}),\ee
where
\begin{equation}\label{eq81}
\begin{array}{c}
  \mathrm{J}=\mathrm{j}\oplus \mathrm{j}\oplus \mathrm{j}\oplus... \\
  \mathrm{J}^{'}=\mathrm{j}^{'}\oplus \mathrm{j}^{'}\oplus \mathrm{j}^{'}\oplus... \\
  \mathrm{J}^{''}=\mathrm{j}^{''}\oplus \mathrm{j}^{''}\oplus \mathrm{j}^{''}\oplus..., \\
\end{array}
\end{equation}, in which the J, $\mathrm{J}^{'}$ and $\mathrm{J}^{''}$ are real skew-symmetric matrices in d-dimensional Hilbert space $\mathcal{H}$ with
\begin{equation}\label{eq81}
\mathrm{j}=\left(
    \begin{array}{cccc}
      0 & 1 & 0 & 0 \\
      -1 & 0& 0 & 0 \\
      0 & 0 & 0 & 1 \\
      0 & 0 & -1 & 0 \\
    \end{array}
  \right),\mathrm{j}^{'}=\left(
                  \begin{array}{cccc}
                    0 & 0 & 0 & 1 \\
                    0 & 0 & 1 & 0 \\
                    0 & -1 & 0 & 0 \\
                    -1 & 0 & 0 & 0 \\
                  \end{array}
                \right),\mathrm{j}^{''}=\left(
                                 \begin{array}{cccc}
                                   0 & 0 & 1 & 0 \\
                                   0 & 0 & 0 & -1 \\
                                   -1 & 0 & 0 & 0 \\
                                   0 & 1 & 0 & 0 \\
                                 \end{array}
                               \right).
\end{equation}
Clearly the ranks of J, $\mathrm{J}^{'}$ and $\mathrm{J}^{''}$ are $4n$, $n=1,2,3,\cdots$ with $d=4n+m$ where $0\leq m\leq3$. The action of $\mathrm{J}$, $\mathrm{J}^{'}$ and $\mathrm{J}^{''}$ on the basis states of the d-dimensional single party subsystem are as
\begin{equation}\label{eq83}
\begin{array}{c}
  \mathrm{J}|4k\rangle=-|4k+1\rangle \\
  \mathrm{J}|4k+1\rangle=|4k\rangle \\
  \mathrm{J}|4k+2\rangle=-|4k+3\rangle \\
  \mathrm{J}|4k+3\rangle=|4k+2\rangle \\
\end{array}, \begin{array}{c}
  \mathrm{J}^{'}|4k\rangle=-|4k+3\rangle \\
  \mathrm{J}^{'}|4k+1\rangle=-|4k+2\rangle \\
  \mathrm{J}^{'}|4k+2\rangle=|4k+1\rangle \\
  \mathrm{J}^{'}|4k+3\rangle=|4k\rangle \\
\end{array},  \begin{array}{c}
  \mathrm{J}^{''}|4k\rangle=-|4k+2\rangle \\
  \mathrm{J}^{''}|4k+1\rangle=|4k+3\rangle \\
  \mathrm{J}^{''}|4k+2\rangle=|4k\rangle \\
  \mathrm{J}^{''}|4k+3\rangle=-|4k+1\rangle \\
\end{array}.
\end{equation}. The expectation values of the operator W with respect to the product states are as
\begin{equation}\label{eq82}
\langle\eta|\otimes\langle\zeta|\mathrm{W}|\eta\rangle\otimes|\zeta\rangle=1-|\langle\zeta|\eta^{\ast}\rangle|^{2}-|\langle\zeta|\mathrm{J}|\eta\rangle|^{2}-|\langle\zeta|\mathrm{J^{'}}|\eta\rangle|^{2}-|\langle\zeta|\mathrm{J^{''}}|\eta\rangle|^{2}.
\end{equation}
It is easy to see that the states $|\eta^{\ast}\rangle$, $J|\eta\rangle$, $\mathrm{J}^{'}|\eta\rangle$ and $\mathrm{J}^{''}|\eta\rangle$ are orthogonal to each other when they are belong to a real single particle Hilbert space. Therefore, the expectation values become positive with respect to all real separable  states. The relation between the operator (\ref{eq12}) and (\ref{eq80}) is
\begin{equation}\label{eq83}
\mathrm{W_{C}}=\mathrm{W}+\mathrm{D}^{T_{A}}_{1}+\mathrm{D}^{T_{A}}_{2},
\end{equation}
where $\mathrm{D}^{T_{A}}_{1}=d(\mathrm{J^{'}}^{T}\otimes
\mathrm{I}_{d})|\psi\rangle\langle\psi|^{T_{A}}(\mathrm{J^{'}}\otimes
\mathrm{I}_{d})$ and $\mathrm{D}^{T_{A}}_{2}=d(\mathrm{J^{''}}^{T}\otimes
\mathrm{I}_{d})|\psi\rangle\langle\psi|^{T_{A}}(\mathrm{J^{''}}\otimes
\mathrm{I}_{d})$. The operators $\mathrm{D}_{1}$ and $\mathrm{D}_{2}$ are positive operators so the PPT entanglement detection power of W is greater than $\mathrm{W_{C}}$. Consider, for example, the following state with $d=4n$ which is defined in a real tensor product Hilbert space $\mathcal{H}_{1}\otimes\mathcal{H}_{2}$ (and keeping in mind that any LOCC must be restricted on the real field) as
$$\rho=a_{0}d|\psi\rangle\langle\psi|+a_{0}\sum_{i=0}^{n-1}[3(|4i,4i\rangle\langle4i,4i|+|4i+1,4i+1\rangle\langle4i+1,4i+1|$$
$$+|4i+2,4i+2\rangle\langle4i+2,4i+2|+|4i+3,4i+3\rangle\langle4i+3,4i+3|)$$
$$-|4i,4i\rangle\langle4i+1,4i+1|-|4i,4i\rangle\langle4i+2,4i+2|-|4i,4i\rangle\langle4i+3,4i+3|$$
$$-|4i+1,4i+1\rangle\langle4i,4i|-|4i+1,4i+1\rangle\langle4i+2,4i+2|-|4i+1,4i+1\rangle\langle4i+3,4i+3|$$
$$-|4i+2,4i+2\rangle\langle4i,4i|-|4i+2,4i+2\rangle\langle4i+1,4i+1|-|4i+2,4i+2\rangle\langle4i+3,4i+3|$$
$$-|4i+3,4i+3\rangle\langle4i,4i|-|4i+3,4i+3\rangle\langle4i+1,4i+1|-|4i+3,4i+3\rangle\langle4i+2,4i+2|]$$
$$+\sum_{i\neq j=0}^{n-1}(a_{4i,4j}|4i,4j\rangle\langle4i,4j|+a_{4i,4j+1}|4i,4j+1\rangle\langle4i,4j+1|+a_{4i,4j+2}|4i,4j+2\rangle\langle4i,4j+2|$$
$$+a_{4i,4j+3}|4i,4j+3\rangle\langle4i,4j+3|+a_{4i+1,4j}|4i+1,4j\rangle\langle4i+1,4j|+a_{4i+1,4j+1}|4i+1,4j+1\rangle\langle4i+1,4j+1|$$
$$+a_{4i+1,4j+2}|4i+1,4j+2\rangle\langle4i+1,4j+2|+a_{4i+1,4j+3}|4i+1,4j+3\rangle\langle4i+1,4j+3|$$
$$+a_{4i+2,4j}|4i+2,4j\rangle\langle4i+2,4j|+a_{4i+2,4j+1}|4i+2,4j+1\rangle\langle4i+2,4j+1|$$
$$+a_{4i+2,4j+2}|4i+2,4j+2\rangle\langle4i+2,4j+2|+a_{4i+2,4j+3}|4i+2,4j+3\rangle\langle4i+2,4j+3|$$
$$+a_{4i+3,4j}|4i+3,4j\rangle\langle4i+3,4j|+a_{4i+3,4j+1}|4i+3,4j+1\rangle\langle4i+3,4j+1|$$
$$+a_{4i+3,4j+2}|4i+3,4j+2\rangle\langle4i+3,4j+2|+a_{4i+3,4j+3}|4i+3,4j+3\rangle\langle4i+3,4j+3|)$$
$$\sum_{i=0}^{n-1}(a_{4i,4i+1}|4i,4i+1\rangle\langle4i,4i+1|+a_{4i,4i+2}|4i,4i+2\rangle\langle4i,4i+2|$$
$$+a_{4i,4i+3}|4i,4i+3\rangle\langle4i,4i+3|+a_{4i+1,4i}|4i+1,4i\rangle\langle4i+1,4i|$$
$$+a_{4i+1,4i+2}|4i+1,4i+2\rangle\langle4i+1,4i+2|+a_{4i+1,4i+3}|4i+1,4i+3\rangle\langle4i+1,4i+3|$$
$$+a_{4i+2,4i}|4i+2,4i\rangle\langle4i+2,4i|+a_{4i+2,4i+1}|4i+2,4i+1\rangle\langle4i+2,4i+1|$$
$$+a_{4i+2,4i+3}|4i+2,4i+3\rangle\langle4i+2,4i+3|+a_{4i+3,4i}|4i+3,4i\rangle\langle4i+3,4i|$$
$$+a_{4i+3,4i+1}|4i+3,4i+1\rangle\langle4i+3,4i+1|+a_{4i+3,4i+2}|4i+3,4i+2\rangle\langle4i+3,4i+2|)$$
$$+a_{0}\sum_{i=0}^{n-1}(|4i+1,4i+2\rangle\langle4i+3,4i|+|4i,4i+3\rangle\langle4i+2,4i+1|$$
\be\label{84}+|4i+3,4i\rangle\langle4i+1,4i+2|+|4i+2,4i+1\rangle\langle4i,4i+3|)\ee
Its positivity and PPT conditions are as
$$a_{0}\geq0,\qquad a_{4i+1,4i+2}a_{4i+3,4i}\geq a_{0}^{2},\qquad a_{4i,4i+3}a_{4i+2,4i+1}\geq a_{0}^{2},$$
\be\label{85}i=0,...,n-1\ee and
$$a_{4i,4j}a_{4j,4i}\geq a_{0}^{2},\qquad a_{4i,4j+1}a_{4j+1,4i}\geq a_{0}^{2},\qquad a_{4i,4j+2}a_{4j+2,4i}\geq a_{0}^{2},$$
$$a_{4i,4j+3}a_{4j+3,4i}\geq a_{0}^{2},\qquad a_{4i+1,4j+1}a_{4j+1,4i+1}\geq a_{0}^{2},\qquad a_{4i+1,4j+2}a_{4j+2,4i+1}\geq a_{0}^{2},$$
$$a_{4i+1,4j+3}a_{4j+3,4i+1}\geq a_{0}^{2},\qquad a_{4i+2,4j+2}a_{4j+2,4i+2}\geq a_{0}^{2},\qquad a_{4i+2,4j+3}a_{4j+3,4i+2}\geq a_{0}^{2},$$
$$a_{4i+3,4j+3}a_{4j+3,4i+3}\geq a_{0}^{2},$$
$$i,j=0,...,n-1,\qquad i\neq j,$$
\be\label{86}a_{4i+1,4i}a_{4i+3,4i+2}\geq a_{0}^{2},\qquad a_{4i,4i+1}a_{4i+2,4i+3}\geq a_{0}^{2},\qquad i=0,...,n-1,\ee respectively.
The expectation value of W with respect to $\rho$, along with the positivity and PPT conditions, satisfies the following inequality
\begin{equation}\label{eq87}
\mathrm{Tr}(\mathrm{W}\rho)\geq-4a_{0}n.
\end{equation}Clearly, by the relation (\ref{eq83}), the PPT entanglement detection power of $\mathrm{W}_{C}$ for this state is weaker than W.
\section{Conclusions}
\par\quad  In this paper, we have constructed new types of EWs, by using real skew-symmetric operators and maximal entangled state
 for bipartite $d\otimes d$ quantum systems analytically. The
proving of the positivity of expectation values of these witnesses
has been done with respect to the separable states very easily. We
have seen that by using various orthogonal transformations on
canonical EWs one can obtain a large number of EWs. It has been
shown that all of the canonical EWs in various ranks are optimal.
When J is full-rank, the canonical EW is optimal nd-EW and for the
cases which is not full-rank ($2n<d$) the corresponding canonical
EW is not optimal nd-EW. In these cases, optimal nd-EW lies in the
$2n\otimes2n$ subspace of $d\otimes d$ Hilbert space. We have also
constructed positive maps corresponding to canonical EWs by
Jamiolkowski isomorphism. On the other hand, we have constructed
the other types of witnesses in $d\otimes d$ Hilbert space
corresponding to the possible partitions of full-rank J. It has
been shown that for a full-rank J there exist p(n) (the number of
partitions of n) number of optimal EWs. Among these witnesses,
there are one optimal d-EW and one optimal nd-EW. The other ones
are composed of optimal nd-EWs and optimal d-EWs. We have also
generalized our approach to the $d_{1}\otimes d_{2}$ ($d_{1}\leq
d_{2}$) quantum systems. We have shown that there exist
$\mathrm{C}^{d_{2}}_{d_{1}}$ distinct possibilities to construct
$\mathrm{EW}$s for a given $d_{1}\otimes d_{2}$ Hilbert space. In
all of the cases where we have discussed, we have emphasized that
the $rank(\mathrm{J})\geq4$ ($rank(\mathrm{J_{1}})\geq4$), because
if we take $rank(\mathrm{J})<4$ ($rank(\mathrm{J_{1}})<4$) then we
can not find out any nd-$\mathrm{EW}$s. Also the idea of canonical EW has been extended to produce other EWs allowed to use only for detecting the entanglement of states which lie in a real tensor product Hilbert space.
\par\quad In each step, we have constructed a class of
PPT states. The expectation values of constructed witnesses with
respect to the corresponding PPT states give out a negative lower
bound. This lower bound lies at the boundary of the entangled PPT
states and entangled states that violate the PPT conditions.
Finally we must mention that the other interesting issues remain
unsolved such as for positive expectation values; we can not
conclude that the corresponding states are separable or not. On
the other hand, by replacing the $\mathrm{W}_{red}$ by the
generalized reduction EW introduced in \cite{jafari6}, the
generalized canonical EWs may be obtained. The work on these
important points is under investigation by these
authors.
\newpage

\vspace{1cm} \setcounter{section}{0}
 \setcounter{equation}{0}
 \renewcommand{\theequation}{A-\arabic{equation}}
{\Large{Appendix A:}}\\
{\bf Proving the lower bound (\ref{eq32})}: \\
\par The expectation value of the EW (\ref{eq23}) with density
matrix $\rho$ (without normalization) in (\ref{eq28}) is given as
$$\mathrm{Tr}(\mathrm{W}_{C}\rho)=-(4n(n-1)+(d-2n)(d+2n-1))a_{0}+\sum_{i\neq j=0}^{n-1}(a_{_{2i,2j}}+a_{_{2i+1,2j+1}}+a_{_{2i+1,2j}}+a_{_{2i,2j+1}})$$
\be\label{A1}-2\sum_{i=0}^{n-1}C_{i}+\sum_{i=2n}^{d-1}\sum_{j=0}^{n-1}(a_{_{2i,j}}+a_{_{j,2i}}+a_{_{2i+1,j}}+a_{_{j,2i+1}})+\sum_{i\neq
j=2n}^{d-1}a_{_{i,j}}\ee Let us rewrite the the PPT condition in
(\ref{eq31}) as

$$a_{_{2i,2j}}\\\\\\\ a_{_{2j,2i}}=\delta^{2}_{2i,2j}a_{0}^{2}\quad ;\quad\delta_{2i,2j}\geq1\quad ;\quad i,j=0,...,n-1\quad;\quad i\neq
j$$
$$a_{_{2i+1,2j+1}}\\\\\\\ a_{_{2j+1,2i+1}}=\delta^{2}_{2i+1,2j+1}a_{0}^{2}\quad ;\quad\delta_{2i+1,2j+1}\geq1\quad ;\quad i,j=0,...,n-1\quad;\quad i\neq
j$$
$$a_{_{2i,2j+1}}\\\\\\\ a_{_{2j+1,2i}}=\delta^{2}_{2i,2j+1}a_{0}^{2}\quad ;\quad\delta_{2i,2j+1}\geq1\quad ;\quad i,j=0,...,n-1\quad;\quad i\neq
j$$
$$a_{_{2i+1,2i}}\\\\\\\ a_{_{2i+2,2i+3}}=\delta^{2}_{2i,2i+2}a_{0}^{2}\quad ,\quad\delta_{2i,2i+2}a_{0}\geq C_{i} \quad ;\quad i=0,...,n-1$$
$$a_{_{2j,i}}\\\\\\\ a_{i,2j}=\delta^{2}_{2j,i}a_{0}^{2}\quad ;\quad\delta_{2j,i}\geq1\quad ;\quad j=0,...,n-1\quad;\quad i=2n,...,d-1$$
$$a_{_{2j+1,i}}\\\\\\\ a_{i,2j+1}=\delta^{2}_{2j+1,i}a_{0}^{2}\quad ;\quad\delta_{2j+1,i}\geq1\quad ;\quad j=0,...,n-1\quad;\quad i=2n,...,d-1$$
\be\label{A2}a_{i,j}\\\\\\\
a_{j,i}=\delta^{2}_{i,j}a_{0}^{2}\quad;\quad\delta_{i,j}\geq1\quad
;\quad i,j=2n,...,d-1\quad;\quad i\neq j\ee

It is clear that $\delta$ is invariant under permutation of its
subscripts. It should be noted that for two variables with
constant product, their summation becomes minimum when they are
equal. The corresponding summation of each product in (\ref{A2})
appears in (\ref{A1}). Hence the minimum value for (\ref{A1}) is
obtained when
$$a_{_{2i,2j}}=a_{_{2j,2i}}=\delta_{2i,2j} a_{0}\quad ;\quad\delta_{2i,2j}=\delta_{2j,2i}\quad ;\quad i,j=0,...,n-1\quad;\quad i\neq
j$$
$$a_{_{2i+1,2j+1}}= a_{_{2j+1,2i+1}}=\delta_{2i+1,2j+1} a_{0}\quad ;\quad\delta_{2i+1,2j+1}=\delta_{2j+1,2i+1}\quad ;\quad i,j=0,...,n-1\quad;\quad i\neq
j$$
$$a_{_{2i,2j+1}}=a_{_{2j+1,2i}}=\delta_{2i,2j+1} a_{0}\quad ;\quad\delta_{2i,2j+1}=\delta_{2j+1,2i}\quad ;\quad i,j=0,...,n-1\quad;\quad i\neq
j$$
$$a_{_{2j,i}}=a_{i,2j}=\delta_{i,2j} a_{0}\quad ;\quad\delta_{i,2j}=\delta_{2j,i}\quad ;\quad j=0,...,n-1\quad;\quad i=2n,...,d-1$$
$$a_{_{2j+1,i}}=a_{i,2j+1}=\delta_{i,2j+1} a_{0}\quad ;\quad\delta_{i,2j+1}=\delta_{2j+1,i}\quad ;\quad j=0,...,n-1\quad;\quad i=2n,...,d-1$$
$$a_{i,j}=a_{j,i}=\delta_{i,j} a_{0}\quad
;\quad\delta_{i,j}=\delta_{j,i}\quad ;\quad
i,j=2n,...,d-1\quad;\quad i\neq j$$ \be\label{A3}
C_{i}=\delta_{2i,2i+2}a_{0}\quad ;\quad i=0,...,n-1\ee Therefore
the equations in (\ref{A3}) satisfy the PPT conditions and ensure
to obtain minimum value for (\ref{A1}). Also the (\ref{A3}) along
with the inequality $C_{i}\leq\delta_{2i+1,2i+3}a_{0}$
($i=0,...,n-1$) gives the positivity conditions for $\rho$. To
check this matter, one can write the first and second line of
(\ref{A3}) for $j=i+1$ and therefore
$a_{_{2i+2,2i}}=\delta_{2i,2i+2}a_{0}$ and
$a_{_{2i+1,2i+3}}=\delta_{2i+1,2i+3}a_{0}$ ($i=0,...,n-1$) so
$a_{_{2i+2,2i}}a_{_{2i+1,2i+3}}=\delta_{2i,2i+2}\delta_{2i+1,2i+3}a_{0}^{2}\geq
C^{2}_{i}$ ($i=0,...,n-1$) which is the same as the positivity
condition for $\rho$ in (\ref{eq30}). The equation (\ref{A1})
becomes
$$\mathrm{Tr}(\mathrm{W}_{C}\rho)=-(4n(n-1)+(d-2n)(d+2n-1))a_{0}+a_{0}\sum_{i\neq j=0}^{n-1}(\delta_{_{2i,2j}}+\delta_{_{2i+1,2j+1}}+\delta_{_{2i+1,2j}}+\delta_{_{2i,2j+1}})$$
\be\label{A4}-a_{0}(\sum_{i=0}^{n-1}\delta_{2i,2i+2}+\delta_{2i+2,2i})+a_{0}\sum_{i=2n}^{d-1}\sum_{j=0}^{n-1}(\delta_{_{2i,j}}+\delta_{_{j,2i}}+\delta_{_{2i+1,j}}+\delta_{_{j,2i+1}})+a_{0}\sum_{i\neq
j=2n}^{d-1}\delta_{_{i,j}}\ee It is clear that we have used the
invariancy of $\delta$ with respect to permutation of its
subscripts. Since $\mathrm{Tr}(\mathrm{W}_{C}\rho)$ is a linear
strictly increasing function of various $\delta$s then its minimum
takes place in lower bounds of various $\delta$s. Thus, when all
of $\delta$s are equal to one the lower bound of (\ref{A4}) is
($-2na_{0}$). Consequently, by considering the normalization
factor $\mathcal{N}$ (\ref{eq29}), which by (\ref{A3}) gets its
minimum value, we obtain
\begin{eqnarray}\label{B5}
\mathrm{Tr}(\mathrm{W}_{C}\rho)\geq\frac{-2n}{d+4n^{2}+(d-2n)(d+2n-1)}\qquad
; \qquad n=2,3,4,\cdots
\end{eqnarray} So we conclude that, the lower bound is obtained in the
boundary of the entangled PPT states and entangled states that
violate the PPT conditions.\\
\vspace{1cm} \setcounter{section}{0}
 \setcounter{equation}{0}
 \renewcommand{\theequation}{B-\arabic{equation}}
{\Large{Appendix B:}}\\
Let us assume that $\mathrm{H}_{2n}\otimes \mathrm{H}_{2n}$ is a subspace of $\mathrm{H}_{d}\otimes \mathrm{H}_{d}$ Hilbert space. $\mathrm{W}_{OPC}$ in (\ref{eq27}), is a Hermitian operator in $\mathrm{H}_{2n}\otimes \mathrm{H}_{2n}$ which is embedded in the $\mathrm{H}_{d}\otimes \mathrm{H}_{d}$ and then its compact form is as
\begin{equation}\label{B1}
\mathrm{W}_{OPC}=\mathrm{I}_{2n}\otimes
\mathrm{I}_{2n}-2n|\psi\rangle\langle\psi|-2n(\mathrm{J}^{T}\otimes
\mathrm{I}_{2n})|\psi\rangle\langle\psi|^{T_{A}}(\mathrm{J}\otimes
\mathrm{I}_{2n})
\end{equation}
where
\begin{equation}\label{B1}
|\psi\rangle=\frac{1}{\sqrt{2n}}\sum_{i=0}^{2n-1}|i,i\rangle
\end{equation}
Now we define $|\eta'\rangle\otimes|\zeta'\rangle\in \mathrm{H}_{2n}\otimes \mathrm{H}_{2n}$ as the projection of products $|\eta\rangle\otimes|\zeta\rangle\in \mathrm{H}_d\otimes \mathrm{H}_d$ so the expectation value of $\mathrm{W}_{OPC}$ with normalized products $|\eta\rangle\otimes|\zeta\rangle$ is
\begin{equation}\label{B3}
\langle\eta|\otimes\langle\zeta|\mathrm{W}_{OPC}|\eta\rangle\otimes|\zeta\rangle=\langle\eta'|\eta'\rangle \langle\zeta'|\zeta'\rangle-|\langle\zeta'|\eta'^{\ast}\rangle|^{2}-|\langle\zeta'|\mathrm{J}|\eta'\rangle|^{2}
\end{equation}
By normalizing the states $|\eta'\rangle$ and $|\zeta'\rangle$, we obtain the next equation
\begin{equation}\label{B4}
\langle\eta|\otimes\langle\zeta|\mathrm{W}_{OPC}|\eta\rangle\otimes|\zeta\rangle=\langle\eta'|\eta'\rangle \langle\zeta'|\zeta'\rangle(1-|\langle\zeta''|\eta''^{\ast}\rangle|^{2}-|\langle\zeta''|\mathrm{J}|\eta''\rangle|^{2})
\end{equation}
where
\begin{equation}\label{B5}
|\eta''\rangle=\frac{1}{\sqrt{\langle\eta'|\eta'\rangle}}|\eta'\rangle \quad,\quad |\zeta''\rangle=\frac{1}{\sqrt{\langle\zeta'|\zeta'\rangle}}|\zeta'\rangle
\end{equation}
The right hand side of the equation (\ref{B4}), apart from the multiplier $\langle\eta'|\eta'\rangle \langle\zeta'|\zeta'\rangle$, is similar to the right hand side of the equation (\ref{eq6}). Since J is full-rank in 2n-dimensional subspace then, by the same argument sketched in section 2, the above mentioned expectation value is zero by the products such as $|\eta''\rangle\otimes(\alpha|\eta''^{\ast}\rangle+\beta J|\eta''\rangle)$ and positive with other separable states so $\mathrm{W}_{OPC}$ is an EW. To further illustrate, if $|\eta\rangle\otimes|\zeta\rangle\in \mathrm{H}_{2n}\otimes \mathrm{H}_{2n}$ then the equation (\ref{B4}) becomes as the following equation
\begin{equation}\label{B6}
\langle\eta|\otimes\langle\zeta|\mathrm{W}_{OPC}|\eta\rangle\otimes|\zeta\rangle=1-|\langle\zeta|\eta^{\ast}\rangle|^{2}-|\langle\zeta|\mathrm{J}|\eta\rangle|^{2}
\end{equation}
which was discussed in section 2. On the other hand, if $|\eta\rangle\otimes|\zeta\rangle\in \mathrm{H}_{d-2n}\otimes \mathrm{H}_{d-2n}$, where $\mathrm{H}_{d-2n}\otimes \mathrm{H}_{d-2n}$ is the complement subspace of $\mathrm{H}_{2n}\otimes \mathrm{H}_{2n}$, then the expectation value of the equation (\ref{B4}) with these products becomes zero because they have no component in $\mathrm{H}_{2n}\otimes \mathrm{H}_{2n}$. Consequently, the expectation value of $\mathrm{W}_{OPC}$ with the following products is always zero,
$$\mathrm{P}_{\mathrm{W}_{OPC}}=\{|\eta\rangle\otimes|\zeta\rangle\in \mathrm{H}_{2n}\otimes \mathrm{H}_{2n} ; |\zeta\rangle=\alpha|\eta^{\ast}\rangle+\beta \mathrm{J}|\eta\rangle\}$$
\be\label{B8}\cup\{|\eta\rangle\otimes|\zeta\rangle\in \mathrm{H}_{d}\otimes \mathrm{H}_{d} ;|\zeta''\rangle= \alpha|\eta''^{\ast}\rangle+\beta J|\eta''\rangle\}\cup\{|\eta\rangle\otimes|\zeta\rangle\in \mathrm{H}_{d-2n}\otimes \mathrm{H}_{d-2n}\}\ee.where $|\eta''\rangle$ and $|\zeta''\rangle$
have been defined in equation (\ref{B5}) and $|\eta'\rangle\otimes|\zeta'\rangle$ is the projection of $|\eta\rangle\otimes|\zeta\rangle$ in to subspace $\mathrm{H}_{2n}\otimes \mathrm{H}_{2n}$ and with other products is positive.

{\Large{Appendix C:}}\\
\par As we considered in the paper, for a given partition ($\mu_{_{1}},\mu_{_{2}},\mu_{_{3}},\cdots,\mu_{_{\nu}}$)
of $n$, the corresponding entanglement witness is given by
$$\mathrm{W_{C}}(\mu_{1},\mu_{2},\mu_{3},\cdots,\mu_{\nu})=\mathcal{W_{C}}(\mu_{1},\mu_{2},\mu_{3},\cdots,\mu_{\nu})+\mathrm{O}^{t_{A}}(\mu_{1})+\mathrm{O}^{t_{A}}(\mu_{1})+\mathrm{O}^{t_{A}}(\mu_{3})+\cdots+\mathrm{O}^{t_{A}}(\mu_{\nu-1})$$
in which
$$\mathcal{W}_{C}(\mu_{1},\mu_{2},\mu_{3},\cdots,\mu_{\nu})=\mathrm{W}_{C}(\mu_{1})\oplus \mathrm{W}_{C}(\mu_{2})\oplus
\mathrm{W}_{C}(\mu_{3})\oplus\cdots\oplus\mathrm{W}_{C}(\mu_{\nu})$$
Each of $\mathrm{W}_{C}(\mu_{\epsilon})$
($\epsilon=1,2,\cdots,\nu$) is canonical $\mathrm{EW}$ on the
$2\mu_{_{\epsilon}}\otimes2\mu_{_{\epsilon}}$ subspace of
$d\otimes d$ Hilbert space such as
$$\mathrm{W_{C}}(\mu _{\epsilon})=\sum_{_{i\neq j=\mu_{1}+\mu_{2}+\cdots+\mu_{_{\epsilon-1}}}}^{\mu_{1}+\mu_{2}+\cdots+\mu_{_{\epsilon}}-1}(|2i,2j\rangle\langle2i,2j|
+|2i,2j+1\rangle\langle2i,2j+1| +|2i+1,2j\rangle\langle2i+1,2j|$$
$$+|2i+1,2j+1\rangle\langle2i+1,2j+1|-|2i,2i\rangle\langle2j,2j|-|2i,2i\rangle\langle2j+1,2j+1|$$
$$-|2i+1,2i+1\rangle\langle2j,2j|-|2i+1,2i+1\rangle\langle2j+1,2j+1|$$
$$-|2i+1,2j\rangle\langle2j+1,2i|+|2i+1,2j+1\rangle\langle2j,2i|$$
$$+|2i,2j\rangle\langle2j+1,2i+1|-|2i,2j+1\rangle\langle2j,2i+1|)$$
As shown in the paper, these witnesses for $\mu_{\epsilon}=1$ are
zero and for the others are optimal nd-$\mathrm{EW}$s.
Consequently the
$\mathcal{W}_{C}(\mu_{1},\mu_{2},\mu_{3},\cdots,\mu_{\nu})$ is
zero or optimal nd-$\mathrm{EW}$ and the operator
$\mathrm{O}^{t_{A}}(\mu_{\epsilon})$ for
($\epsilon=1,2,3,\cdots\nu-1$) is written as below
$$\mathrm{O}^{t_{A}}(\mu _{\epsilon})=\sum_{_{i=\mu_{1}+\mu_{2}+\cdots+\mu_{_{\epsilon-1}}}}^{\mu_{1}+\mu_{2}+\cdots+\mu_{_{\epsilon}}-1}\sum_{_{j=\mu_{1}+\mu_{2}+\cdots+\mu_{_{\epsilon}}}}^{\frac{d}{2}-1}(|2i,2j\rangle\langle2i,2j|
+|2i,2j+1\rangle\langle2i,2j+1| +|2i+1,2j\rangle\langle2i+1,2j|$$
$$+|2i+1,2j+1\rangle\langle2i+1,2j+1|-|2i,2i\rangle\langle2j,2j|-|2i,2i\rangle\langle2j+1,2j+1|$$
$$-|2i+1,2i+1\rangle\langle2j,2j|-|2i+1,2i+1\rangle\langle2j+1,2j+1|$$
$$|2j,2i\rangle\langle2j,2i|+|2j,2i+1\rangle\langle2j,2i+1|+|2j+1,2i\rangle\langle2j+1,2i|$$
$$+|2j+1,2i+1\rangle\langle2j+1,2i+1|-|2j,2j\rangle\langle2i,2i|-|2j,2j\rangle\langle2i+1,2i+1|$$
$$-|2j+1,2j+1\rangle\langle2i,2i|-|2j+1,2j+1\rangle\langle2i+1,2i+1|)$$
It is clear that the $\mathrm{O}(\mu _{\epsilon})$ is positive
operator so the $\mathrm{O}^{t_{A}}(\mu _{\epsilon})$ is optimal
d-$\mathrm{EW}$.\\

{\Large{Appendix D:}}\\
\par In this appendix we give the positivity and $\mathrm{PPT}$ conditions for the following $\mathrm{PPT}$ state which
corresponds to the partition
($\mu_{_{1}},\mu_{_{2}},\mu_{_{3}},\cdots,\mu_{_{\nu}}$) of
$n=\frac{d}{2}$.
$$\rho(\mu_{1},\mu_{2},\mu_{3},\cdots,\mu_{\nu})$$
$$=\frac{1}{\mathcal{N}(\mu_{1},\mu_{2},\mu_{3},\cdots,\mu_{\nu})}(\varrho(\mu_{1},\mu_{2},\mu_{3},\cdots,\mu_{\nu})+\sigma(\mu_{1})+\sigma(\mu_{1})+\sigma(\mu_{3})+\cdots+\sigma(\mu_{\nu-1}))$$

where
$$\varrho(\mu_{1},\mu_{2},\mu_{3},\cdots,\mu_{\nu})=\rho(\mu_{1})\oplus \rho(\mu_{2})\oplus\rho(\mu_{3})\oplus\cdots \oplus \rho(\mu_{\nu})$$
Each of the $\rho(\mu_{\epsilon})$ with
($\epsilon=1,2,3,\cdots,\nu$) and $\sigma(\mu_{\epsilon})$ with
($\epsilon=1,2,3,\cdots,\nu-1$) is given as below
$$\rho(\mu_{\epsilon})=a_{0}|\psi(\mu_{\epsilon})\rangle\langle\psi(\mu_{\epsilon})|$$
$$+a_{0}\sum_{i=\mu_{_{1}}+\mu_{_{2}}+\mu_{_{3}}+\cdots+\mu_{_{\epsilon-1}}}^{\mu_{_{1}}+\mu_{_{2}}+\mu_{_{3}}+\cdots+\mu_{_{\epsilon}}-1}
(|2i,2i\rangle\langle2i,2i|+|2i+1,2i+1\rangle\langle2i+1,2i+1|$$
$$-|2i,2i\rangle\langle2i+1,2i+1|-|2i+1,2i+1\rangle\langle2i,2i|)$$
$$+\sum_{i=\mu_{_{1}}+\mu_{_{2}}+\mu_{_{3}}+\cdots+\mu_{_{\epsilon-1}}}^{\mu_{_{1}}+\mu_{_{2}}+\mu_{_{3}}+\cdots+\mu_{_{\epsilon}}-1}(a_{2i+2,2i}|2i+2,2i\rangle\langle2i+2,2i|+a_{2i+1,2i+3}|2i+1,2i+3\rangle\langle2i+1,2i+3|$$
$$-C_{i}(|2i+2,2i\rangle\langle2i+1,2i+3|+|2i+1,2i+3\rangle\langle2i+2,2i|))$$
$$+\sum_{i\neq j=\mu_{_{1}}+\mu_{_{2}}+\mu_{_{3}}+\cdots+\mu_{_{\epsilon-1}},i-j\neq1,j-i\neq\mu_{_{\epsilon}}-1}^{\mu_{_{1}}+\mu_{_{2}}+\mu_{_{3}}+\cdots+\mu_{_{\epsilon}}-1}a_{2i,2j}|2i,2j\rangle\langle2i,2j|$$
$$+\sum_{i\neq j=\mu_{_{1}}+\mu_{_{2}}+\mu_{_{3}}+\cdots+\mu_{_{\epsilon-1}},j-i\neq1,i-j\neq\mu_{_{\epsilon}}-1}^{\mu_{_{1}}+\mu_{_{2}}+\mu_{_{3}}+\cdots+\mu_{_{\epsilon}}-1}a_{2i+1,2j+1}|2i+1,2j+1\rangle\langle2i+1,2j+1|$$
$$+\sum_{i,j=\mu_{_{1}}+\mu_{_{2}}+\mu_{_{3}}+\cdots+\mu_{_{\epsilon-1}}}^{\mu_{_{1}}+\mu_{_{2}}+\mu_{_{3}}+\cdots+\mu_{_{\epsilon}}-1}(a_{2i,2j+1}|2i,2j+1\rangle\langle2i,2j+1|+a_{2i+1,2j}|2i+1,2j\rangle\langle2i+1,2j|)$$
in which
$$|\psi(\mu_{\epsilon})\rangle=\frac{1}{\sqrt{2\mu_{\epsilon}}}\sum_{i=1}^{2\mu_{\epsilon}-1}|i,i\rangle$$
is the maximal entangled state in
$2\mu_{\epsilon}\otimes2\mu_{\epsilon}$ subspace.
 and
$$\sigma(\mu _{\epsilon})=\sum_{_{i=\mu_{_{1}}+\mu_{_{2}}+\mu_{_{3}}+\cdots+\mu_{_{\epsilon-1}}}}^{\mu_{_{1}}+\mu_{_{2}}+\mu_{_{3}}+\cdots+\mu_{_{\epsilon}}-1}\sum_{_{j=\mu_{_{1}}+\mu_{_{2}}+\mu_{_{3}}+\cdots+\mu_{_{\epsilon}}}}^{\frac{d}{2}-1}(a_{_{2i,2j}}|2i,2j\rangle\langle2i,2j|
+a_{_{2i,2j+1}}|2i,2j+1\rangle\langle2i,2j+1|$$
$$+a_{_{2i+1,2j}}|2i+1,2j\rangle\langle2i+1,2j|+a_{_{2i+1,2j+1}}|2i+1,2j+1\rangle\langle2i+1,2j+1|$$
$$+a_{0}(|2i,2i\rangle\langle2j,2j|+|2i,2i\rangle\langle2j+1,2j+1|$$
$$+|2i+1,2i+1\rangle\langle2j,2j|+|2i+1,2i+1\rangle\langle2j+1,2j+1|)$$
$$+a_{_{2j,2i}}|2j,2i\rangle\langle2j,2i|+a_{_{2j,2i+1}}|2j,2i+1\rangle\langle2j,2i+1|$$
$$+a_{_{2j+1,2i}}|2j+1,2i\rangle\langle2j+1,2i|+a_{_{2j+1,2i+1}}|2j+1,2i+1\rangle\langle2j+1,2i+1|$$
$$+a_{0}(|2j,2j\rangle\langle2i,2i|+|2j,2j\rangle\langle2i+1,2i+1|$$
$$+|2j+1,2j+1\rangle\langle2i,2i|+|2j+1,2j+1\rangle\langle2i+1,2i+1|))$$
where
$$\mathcal{N}(\mu_{1},\mu_{2},\mu_{3},\cdots,\mu_{\nu})=\sum_{\epsilon=1}^{\nu}\mathcal{N}(\mu_{\epsilon})+\sum_{\epsilon=1}^{\nu-1}n(\mu_{\epsilon})$$
with
$$\mathcal{N}(\mu_{_{\epsilon}})=4a_{0}\mu_{_{\epsilon}}+\sum_{i\neq j=\mu_{_{1}}+\mu_{_{2}}+\mu_{_{3}}+\cdots+\mu_{_{\epsilon-1}}}^{\mu_{_{1}}+\mu_{_{2}}+\mu_{_{3}}+\cdots+\mu_{_{\epsilon}}-1}(a_{_{2i,2j}}+a_{_{2i+1,2j+1}})+\sum_{i,j=\mu_{_{1}}+\mu_{_{2}}+\mu_{_{3}}+\cdots+\mu_{_{\epsilon-1}}}^{\mu_{_{1}}+\mu_{_{2}}+\mu_{_{3}}+\cdots+\mu_{_{\epsilon}}-1}(a_{_{2i,2j+1}}+a_{_{2i+1,2j}})$$
and
$$n(\mu_{_{\epsilon}})=\sum_{i=\mu_{_{1}}+\mu_{_{2}}+\mu_{_{3}}+\cdots+\mu_{_{\epsilon-1}}}^{\mu_{_{1}}+\mu_{_{2}}+\mu_{_{3}}+\cdots+\mu_{_{\epsilon}}-1}\sum_{j=\mu_{_{1}}+\mu_{_{2}}+\mu_{_{3}}+\cdots+\mu_{_{\epsilon}}}^{\frac{d}{2}-1}(a_{_{2i,2j}}+a_{_{2i,2j+1}}+a_{_{2i+1,2j}}+a_{_{2i+1,2j+1}}$$
$$+a_{_{2j,2i}}+a_{_{2j,2i+1}}+a_{_{2j+1,2i}}+a_{_{2j+1,2i+1}})$$
\textbf{The Positivity Conditions:}\\
The following conditions must be satisfied for each
$\epsilon=1,2,3,\cdots,\nu$\\ \textbf{1)}$$a_{0}\geq0$$
\textbf{2)}$$a_{_{2i,2j}}\geq0 \qquad for \qquad
i,j=\mu_{1}+\mu_{2}+\cdots+\mu_{\epsilon-1},\cdots,\mu_{1}+\mu_{2}+\cdots+\mu_{\epsilon}-1$$
$$with \qquad i\neq j \qquad ; \qquad i-j\neq1 \qquad ; \qquad j-i\neq\mu_{\epsilon}-1$$
\textbf{3)}$$a_{_{2i+1,2j+1}}\geq0 \qquad for \qquad i,j=\mu_{1}+\mu_{2}+\cdots+\mu_{\epsilon-1},\cdots,\mu_{1}+\mu_{2}+\cdots+\mu_{\epsilon}-1$$
$$with \qquad i\neq j \qquad ; \qquad j-i\neq1 \qquad ; \qquad i-j\neq\mu_{\epsilon}-1$$
\textbf{4)}$$a_{_{2i,2j+1}}\geq0 \quad ; \quad a_{_{2i+1,2j}}\geq0 \quad for \quad i,j=\mu_{1}+\mu_{2}+\cdots+\mu_{\epsilon-1},\cdots,\mu_{1}+\mu_{2}+\cdots+\mu_{\epsilon}-1$$
\textbf{5)}$$a_{_{2i+2,2i}}a_{_{2i+1,2i+3}}\geq C^{2}_{i} \qquad for \qquad i=\mu_{1}+\mu_{2}+\cdots+\mu_{\epsilon-1},\cdots,\mu_{1}+\mu_{2}+\cdots+\mu_{\epsilon}-1$$
\textbf{6)}$$a_{_{2i,2j}}\geq0 \qquad ; \qquad a_{_{2i,2j+1}}\geq0 \qquad ;\qquad a_{_{2i+1,2j}}\geq0 \qquad ;\qquad a_{_{2i+1,2j+1}}\geq0$$
$$a_{_{2j,2i}}\geq0 \qquad ; \qquad a_{_{2j,2i+1}}\geq0 \qquad ;\qquad a_{_{2j+1,2i}}\geq0 \qquad ;\qquad a_{_{2j+1,2i+1}}\geq0$$
$$for\qquad i=\mu_{1}+\mu_{2}+\cdots+\mu_{\epsilon-1},\cdots,\mu_{1}+\mu_{2}+\cdots+\mu_{\epsilon}-1 \qquad and\qquad j=\mu_{1}+\mu_{2}+\cdots+\mu_{\epsilon},\cdots,\frac{d}{2}-1$$
Therefore these inequalities ensure that
$\rho(\mu_{1},\mu_{2},\mu_{3},\cdots,\mu_{\nu})$ and
$\rho(\mu_{\epsilon})$ ($\epsilon=1,...,\nu$) are positive
operators.

\textbf{The PPT Conditions:}\\
\textbf{1)}$$a_{_{2i,2j}}a_{_{2j,2i}}\geq a^{2}_{0} \qquad ;
\qquad a_{_{2i+1,2j+1}}a_{_{2j+1,2i+1}}\geq a^{2}_{0}\qquad
;\qquad a_{_{2i,2j+1}}a_{_{2j+1,2i}}\geq a^{2}_{0}$$
$$for \qquad i,j=\mu_{1}+\mu_{2}+\cdots+\mu_{\epsilon-1},\cdots,\mu_{1}+\mu_{2}+\cdots+\mu_{\epsilon}-1\qquad with \qquad i\neq j$$
\textbf{2)}$$a_{_{2i+1,2i}}a_{_{2i+2,2i+3}}\geq C^{2}_{i} \qquad
for \qquad
i=\mu_{1}+\mu_{2}+\cdots+\mu_{\epsilon-1},\cdots,\mu_{1}+\mu_{2}+\cdots+\mu_{\epsilon}-1$$
\textbf{3)}$$a_{_{2i,2j}}a_{_{2j,2i}}\geq a^{2}_{0} \quad ; \quad
a_{_{2i,2j+1}}a_{_{2j+1,2i}}\geq a^{2}_{0} \quad ;\quad
a_{_{2i+1,2j}}a_{_{2j,2i+1}}\geq a^{2}_{0} \quad ;\quad
a_{_{2i+1,2j+1}}a_{_{2j+1,2i+1}}\geq a^{2}_{0}$$
$$for\qquad i=\mu_{1}+\mu_{2}+\cdots+\mu_{\epsilon-1},\cdots,\mu_{1}+\mu_{2}+\cdots+\mu_{\epsilon}-1 \qquad and\qquad j=\mu_{1}+\mu_{2}+\cdots+\mu_{\epsilon},\cdots,\frac{d}{2}-1$$
And these inequalities give the PPT conditions for
$\rho(\mu_{1},\mu_{2},\mu_{3},\cdots,\mu_{\nu})$ and
$\rho(\mu_{\epsilon})$ ($\epsilon=1,...,\nu$).

\end{document}